\documentclass[aps,floatfix,preprint,preprintnumbers,nofootinbib,superscriptaddress,natbib]{revtex4}

\usepackage{graphicx,float}
\usepackage{epsfig}		
\usepackage{dcolumn}
\usepackage{bm}
\usepackage{subfigure}

\usepackage[all]{xy}
\usepackage{amsmath,upgreek}
\usepackage{amssymb}

\usepackage{pdfpages}
\usepackage{color,soul}
\usepackage{graphicx,epstopdf}
\usepackage[colorlinks=true,
 linkcolor=red,
 urlcolor=purple,
 citecolor=blue,hyperindex]{hyperref}
\def\0{\mbox{\tiny $0$}}
\def\1{\mbox{\tiny $1$}}
\def\2{\mbox{\tiny $2$}}
\def\3{\mbox{\tiny $3$}}
\def\4{\mbox{\tiny $4$}}
\def\5{\mbox{\tiny $5$}}
\def\6{\mbox{\tiny $6$}}
\def\7{\mbox{\tiny $7$}}
\def\8{\mbox{\tiny $8$}}
\def\9{\mbox{\tiny $9$}}

\def\f14{\mbox{\tiny $\frac{1}{4}$}}

\begin{document}

\title{Extended Weyl-Wigner phase-space framework for non-linear systems: typical and modified prey-predator-like dynamics}

\author{Alex E. Bernardini}
\email{alexeb@ufscar.br}
\affiliation{~Departamento de F\'{\i}sica, Universidade Federal de S\~ao Carlos, PO Box 676, 13565-905, S\~ao Carlos, SP, Brasil.}
\author{O. Bertolami}
\email{orfeu.bertolami@fc.up.pt}
\altaffiliation[Also at~]{Centro de F\'isica do Porto, Rua do Campo Alegre 687, 4169-007, Porto, Portugal.} 
\affiliation{Departamento de F\'isica e Astronomia, Faculdade de Ci\^{e}ncias da Universidade do Porto, Rua do Campo Alegre 687, 4169-007, Porto, Portugal.}
\date{\today}

\begin{abstract}
The extension of the phase-space Weyl-Wigner quantum mechanics to the subset of Hamiltonians in the form of $H(q,\,p) = {K}(p) + {V}(q)$ (with $K(p)$ replacing single $p^2$ contributions) is revisited.
Deviations from classical and stationary profiles are identified in terms of Wigner functions and Wigner currents for Gaussian and gamma/Laplacian distribution ensembles. The procedure is successful in accounting for the exact pattern of quantum fluctuations when compared with the classical phase-space pattern.
General results are then specialized to some specific Hamiltonians revealing non-linear dynamics, and suggest a novel algorithm to treat quantum modifications mapped by Wigner currents. Our analysis shows that the framework encompasses, for instance, the quantized prey-predator-like scenarios subjected to statistical constraints.
\end{abstract}


\date{\today}
\maketitle

\section{Introduction}

The Weyl-Wigner (WW) formalism of quantum mechanics (QM) \cite{Wigner,Case,Weyl} comprises a phase-space quantization scheme which connects phase-space trajectories to operators in the Hilbert space \cite{Abr65,Sch81,Par88}.
This extended view allows for addressesing a plethora of paradigmatic issues which includes the wave-function collapse \cite{Zurek02,Bernardini13A,Leal2019}, the non-commutative QM \cite{Catarina,Bernardini13B,PhysicaA}, generalizations of the correspondence between uncertainty relations and quantum observables \cite{Catarina001,Stein,Bernardini13C,Bernardini13E}, and the close relation between quantum and classical statistical mechanics.

The bridge between operator methods and path integral techniques \cite{Abr65,Sch81,Par88} is encoded by a Weyl transform operation over a generic quantum operator, $\hat{O}$, as defined by
\begin{equation}
O^W(q,\, p)\label{Wigner111}
= 2\hspace{-.15cm} \int^{+\infty}_{-\infty} \hspace{-.35cm}ds\,\exp{\left[2\,i \,p\, s/\hbar\right]}\,\langle q - s | \hat{O} | q + s \rangle=2\hspace{-.15cm} \int^{+\infty}_{-\infty} \hspace{-.35cm} dr \,\exp{\left[-2\, i \,q\, r/\hbar\right]}\,\langle p - r | \hat{O} | p + r\rangle,
\end{equation} 
which, for $2\pi\hbar\,\hat{O}$ identified as the quantum mechanical density matrix operator, $\hat{\rho} = |\psi \rangle \langle \psi|$ (with $\hbar$ denoting the reduced Planck constant),
results in the Wigner phase-space {\em quasi}-distribution function \cite{JCAP18,NossoPaper,Meu2018,Bernardini2020-02} written as
\begin{equation}
W(q,\, p) = (\pi\hbar)^{-1} 
\int^{+\infty}_{-\infty} \hspace{-.35cm}ds\,\exp{\left[2\, i \, p \,s/\hbar\right]}\,
\psi(q - s)\,\psi^{\ast}(q + s),\label{Wigner222}
\end{equation}
i.e. the Weyl transform of the density matrix, which can also be seen as the Fourier transform of the off-diagonal elements of $\hat{\rho}$.

The consistent probability distribution interpretation is identified by the normalization constraint over $\hat{\rho}$, i.e. $Tr_{\{q,p\}}[\hat{\rho}]=1$, which results from straightforward integrations of position and momentum marginal distributions recovered from Eq.~\eqref{Wigner222} as
\begin{equation}
\vert \psi(q)\vert^2 = \int^{+\infty}_{-\infty} \hspace{-.35cm}dp\,W(q,\, p)
\qquad
\leftrightarrow
\qquad
\vert \varphi(p)\vert^2 = \int^{+\infty}_{-\infty} \hspace{-.35cm}dq\,W(q,\, p),
\end{equation}
respectively, such that the Fourier transform relating position and momentum wave functions,
\begin{equation}
 \varphi(p)=
(2\pi\hbar)^{-1/2}\int^{+\infty}_{-\infty} \hspace{-.35cm} dq\,\exp{\left[i \, p \,q/\hbar\right]}\,
\psi(q),
\end{equation}
suppresses the coexistence of positive-definiteness of simultaneous position and momentum Wigner probability distributions.
These are consistent with the Heisenberg-Weyl algebra, which at $1$-dim is driven by the position-momentum non-commutative relation, $[\hat{q},\,\hat{p}] = i\hbar$, both identified as intrinsic Hilbert space features of the Schr\"odinger QM.
In fact, averaged values of quantum observables, $\hat{O}$, are evaluated through an overlap integral over the infinite volume described by the phase-space coordinates, $q$ and $p$, as \cite{Wigner,Case}
\begin{equation}
 \langle O \rangle = 
\int^{+\infty}_{-\infty} \hspace{-.35cm}dp\int^{+\infty}_{-\infty} \hspace{-.35cm} {dq}\,\,W(q,\, p)\,{O^W}(q,\, p), \label{eqfive}
\end{equation}
which corresponds to the trace of the product between $\hat{\rho}$ and $\hat{O}$, $Tr_{\{q,p\}}\left[\hat{\rho}\hat{O}\right]$.
The replacement of ${O^W}(q,\, p)$, for instance, by $W(q,\, p)$, into Eq.~\eqref{eqfive}, leads to the quantum purity computed through an analogue of the trace operation, $Tr_{\{q,p\}}[\hat{\rho}^2]$, read as
\begin{equation}
Tr_{\{q,p\}}[\hat{\rho}^2] = 2\pi\hbar\int^{+\infty}_{-\infty}\hspace{-.35cm}dp\int^{+\infty}_{-\infty} \hspace{-.35cm} {dq}\,\,\,W(q,\, p)^2,
\label{eqpureza}
\end{equation}
which satisfies the pure state constraint, $Tr_{\{q,p\}}[\hat{\rho}^2] = Tr_{\{q,p\}}[\hat{\rho}] = 1$.

More interestingly, the WW formalism can also be implemented through a probability flux continuity equation, through which the fluid analogy associated with some mathematical manipulations can resolve the dynamics of quantum ensembles, as opposed to quantum states.
In this case, one can say that the dynamics is described by a {\em Hamiltonian constraint}, as opposed to a Hamiltonian function.
The phase-phase information flow \cite{Steuernagel3,NossoPaper,Meu2018}, in such a context, encodes all the information provided by a quantum density matrix operator, and classical-quantum limits of exact quantum solutions can be tested by means of probability distributions and information quantifiers, all obtained from the WW formalism. 

In this manuscript, the range of the formalism will be broadened to consider the Wigner flow framework for Hamiltonian systems that can be cast in the form of $H^{W}(q,\,p) = K(p) + V(q)$, with $K(p)$ and $V(q)$ corresponding to arbitrary functions of momentum and position, respectively, for which the probability currents, and the Hamiltonian related (quantum) information quantifiers can be obtained.
Our results follow from considering 
Gaussian \cite{Novo21A,Novo21B,Novo21B2} and gamma/Laplacian distribution phase-space ensembles in order to obtain a novel algorithm to treat quantum modifications mapped by Wigner currents such that classical-quantum limits encompassed by exact quantum solutions can be tested by means of phase-space information quantifiers. As a final proposal, some analytical results are specialized to Gaussian and gamma/Laplacian distribution ensembles when driven by some exotic Hamiltonians which include, for instance, those ones describing prey-predator-like dynamics.

In such a context, it is worth mentioning that quantum-based frameworks \cite{0001,0002,0003} for modeling competitive microscopic systems, as well as for explaining self-organizing complex hierarchical structures \cite{0004}, have already been investigated in the framework of quantum paradigms, either under theoretical \cite{Novo21B,Novo21B2,Novo222} or under phenomenological perspectives. This includes, in the latter approach, the investigation of prey-predator population oscillations, chaos induced by competition dynamics, and molecular programing algorithms for symbiotic synchronization \cite{PP00,PP01,PP02,PP03,PP04}.
Likewise, through the extended WW formalism, classical and quantum descriptions of the problem of population oscillation dynamics have already been connected by Hamiltonian equations of motion when they are convolved by suitable phase-space Wigner statistical distributions \cite{Novo21B,Novo21B2,Novo222}. It is expected that the inclusion of gamma/Laplacian distribution ensembles in the analysis of typical and modified prey-predator-like dynamics may affect the so-called (prey and predator) dominating phases emerging from quantum effects.

The outline of the manuscript is thus as follows.
In Sec.~II, after reporting about the Wigner flow properties, in correspondence with Schr\"odinger QM, the WW framework extended to a subset of Hamiltonians in the form of $H(q,\,p) = {K}(p) + {V}(q)$ is re-obtained \cite{Novo21A}, and stationarity and Liouvillianity quantifiers are identified.
A revision of results for Gaussian ensembles is provided in Sec.~III.
The novel quantitative analysis considering gamma/Laplacian distribution ensembles is provided in Sec.~IV, such that Wigner currents and associated information quantifiers can be computed in terms of a convergent infinite series expansion over $\hbar^{2n}$.
In Sec.~V, the Lotka-Volterra (LV) Hamiltonian \cite{LV1,LV2} and some related modifications are discussed in order to investigate the prey-predator-like-dynamics and their corresponding classical-to-quantum correspondence. In particular, the persistence of prey and predator dominating phases are investigated through suitable modifications on the LV Hamiltonian.
In case of gamma/Laplacian distributions, quantifiers of stationarity and Liouvillianity are all explicitly obtained in terms of analytical expressions for the Wigner currents.
Our conclusions and the outlook for further research are presented in Sec.~VI.

\section{Extended Weyl-Wigner framework}

The WW phase-space formulation encompasses all QM formalisms \cite{Ballentine,Case,Meu2018}, with the Wigner function dynamical properties connected to the Hamiltonian dynamics
by means of a vector flux \cite{Steuernagel3,NossoPaper,Meu2018}, $\mbox{\boldmath ${J}$}(q,\,p;\,t)$, through the continuity equation \cite{Case,Ballentine,Steuernagel3,NossoPaper,Meu2018},
\begin{equation}
{\partial_t W} +\mbox{\boldmath $\nabla$}_{\{q,p\}} \cdot \mbox{\boldmath ${J}$} =0,
\label{alexquaz51}
\end{equation}
with $\mbox{\boldmath $\nabla$}_{\{q,p\}} \cdot \mbox{\boldmath ${J}$} = {\partial_q J_q}+{\partial_p J_p}$, where $\mbox{\boldmath ${J}$}(q,\,p;\,t)$ has been decomposed into the phase-space coordinate directions as $\mbox{\boldmath ${J}$} ={J}_q\,\hat{q} + {J}_p\,\hat{p}$, and the shortened notation for partial derivatives was set as $\partial_a \equiv \partial/\partial a$.

\subsection{Wigner currents in correspondence with Schr\"odinger QM}

For a non-relativistic Hamiltonian operator, 
\begin{equation}
{H}(\hat{Q},\,\hat{P}) = \frac{\hat{P}^2}{2m} + V(\hat{Q}),
\end{equation}
from which the Weyl transform yields 
\begin{equation}
H^{W}(q,\, p) = \frac{{p}^2}{2m} + V(q),
\end{equation}
Wigner currents are written as \cite{Case,Ballentine,Steuernagel3,NossoPaper}
\begin{equation}
J_q(q,\,p;\,t)= \frac{p}{m}\,W(q,\,p;\,t), \label{alexquaz500BB}
\end{equation}
and
\begin{equation}
J_p(q,\,p;\,t) = -\sum_{\eta=0}^{\infty} \left(\frac{i\,\hbar}{2}\right)^{2\eta}\frac{1}{(2\eta+1)!} \, \left[\partial_q^{2\eta+1}V(q)\right]\,\partial_p ^{2\eta}W(q,\,p;\,t),
\label{alexquaz500}
\end{equation}
where, in the above series expansion, contributions from $\eta \geq 1$ describe quantum corrections (eventually coupled to non-linear contributions from $\partial_q^{2\eta+1}V(q)$) which distort phase-space classical trajectories. Of course, the suppression of the $\eta \geq 1$ contributions results in a classical ($\mathcal{C}$) Hamiltonian description in terms of Wigner currents:
\begin{equation}
J^{\mathcal{C}}_q(q,\,p;\,t)= +({\partial_p H^{W}})\,W(q,\,p;\,t), \label{alexquaz500BB2}
\end{equation}
and
\begin{equation}
J^{\mathcal{C}}_p(q,\,p;\,t) = -({\partial_q H^{W}})\,W(q,\,p;\,t),
\label{alexquaz500CC2}
\end{equation}
which, once substituted into Eq.~\eqref{alexquaz51}, result in the classical Liouville equation.
In this case, the classical phase-space velocity is identified by $\mathbf{v}_{(\mathcal{C})} = (\dot{q},\,\dot{p})\equiv ({\partial_p H^{W}},\,-{\partial_q H^{W}})$, with $\mbox{\boldmath $\nabla$}_{\{q,p\}}\cdot \mathbf{v}_{(\mathcal{C})}= \partial_q \dot{q} + \partial_p\dot{p} = 0$, with {\em dots} denoting the time derivative, $d/dt$.


\subsection{Wigner currents in the extended framework}

Given such universal features of the WW formalism, extending the phase-space Schr\"odinger analogue formulation of QM to Hamiltonian systems which reveal a non-linear dynamics could be thought as a natural issue. 
However, in comparison with the classical Hamiltonian formulation, QM states are based on Hilbert spaces and operators implemented through the Schr\"odinger equation, whereas classical mechanics is geometrically defined on symplectic manifolds, whose dynamic trajectories are described by Hamilton's equations.
For more general (Weyl transformed) Hamiltonians generically described by
\begin{equation}
H^{W}(q,\,p) = K(p) + V(q),
\label{nlh}
\end{equation}
where $K(p)$ replaces $p^2$, implementing the {\em Hamiltonian function} through an {\em eigen}system, $H^{W}\, \psi_n = E_n\, \psi_n$, is sometimes unfeasible, even numerically.

To approach such an issue, our departing point is the Von Neumann equation for the state density operator, $\hat{\rho} = \vert \psi \rangle \langle \psi\vert$, written in the form of \cite{Novo21A,Ballentine}
\begin{equation}
\partial_t\hat{\rho} = i\hbar^{-1} \left[\hat{\rho}, \, H \right] \equiv{\partial_t^{^{(K)}}\hat{\rho}} ~ + ~{\partial_t^{^{(V)}}\hat{\rho}},\quad \mbox{with} ~~ {\partial_t^{^{( \mathcal{A})}}\hat{\rho}} = i\hbar^{-1} \left[\hat{\rho}, \, \mathcal{A}\right],
\label{dens}
\end{equation}
which can then be separately evaluated in momentum and position representations, for $\mathcal{A} \equiv K(\hat{P}), \,V(\hat{Q})$.
Hence, using the Wigner function properties from Eq.~\eqref{Wigner222} to transform each contribution into its respective Wigner representation (cf. Ref.~\cite{Ballentine} for non-relativistic QM, and Ref.~\cite{Novo21A} for the extended framework), one has, firstly, in the momentum representation,
\begin{eqnarray}
\partial^{^{(K)}}_t\langle p \vert {\rho} \vert p'\rangle &=& i \hbar^{-1}\langle p \vert {\rho} \vert p'\rangle
\,\left(K(p') - K(p)\right),
\end{eqnarray}
which expands to
\begin{eqnarray}
\partial^{^{(K)}}_t W(q,\,p;\,t) &=& i \hbar^{-1} (\pi\hbar)^{-1}
\int^{+\infty}_{-\infty} \hspace{-.35cm}dr\,
\rho^{W,\varphi}_{(p-r;\,p+r)}
\exp{\left[-2\, i \, q \,r/\hbar\right]}
\,\left[K(p+r) - K(p-r)\right],\nonumber
\label{Wigner222BB}
\end{eqnarray}
where $\rho^{W,\varphi}_{(p-r;\,p+r)} \equiv \langle p-r \vert {\rho} \vert p+r\rangle$ takes the place of $\varphi(p- r)\,\varphi^{\ast}(p+ r)$, and secondly, in the position representation,
\begin{eqnarray}
\partial^{^{(V)}}_t\langle q \vert {\rho} \vert q'\rangle &=& i \hbar^{-1}\langle q \vert {\rho} \vert q'\rangle
\,\left[V(q') - V(q)\right],
\end{eqnarray}
which expands to
\begin{eqnarray}
 \partial^{^{(V)}}_t W(q,\,p;\,t) &=& i \hbar^{-1} (\pi\hbar)^{-1}
\int^{+\infty}_{-\infty} \hspace{-.35cm}ds\,
\rho^{W,\psi}_{(q-s;\,q+s)}
\exp{\left[2\, i \, p \,s/\hbar\right]}
\,\left(V(q+s) - V(q-s)\right),\nonumber
\label{Wigner222CC}
\end{eqnarray} 
where $\rho^{W,\psi}_{(q-s;\,q+s)} \equiv \langle q - s \vert {\rho} \vert q + s\rangle$ takes the place of $\psi(q - s)\,\psi^{\ast}(q + s)$.

Now, simple manipulations yield
\begin{equation}
K(p+r) - K(p-r) = 2\sum_{\eta=0}^{\infty}\frac{r^{2\eta+1}}{(2\eta+1)!} \,\partial_p^{2\eta+1}K(p),
\label{alexquaz500BB}
\end{equation}
and
\begin{equation}
V(q+s) - V(q-s) = 2\sum_{\eta=0}^{\infty}\frac{s^{2\eta+1}}{(2\eta+1)!} \,\partial_q^{2\eta+1}
V(q),
\label{alexquaz500CC}
\end{equation}
with $r$ and $s$ identified by $+i(\hbar/2)\, \partial_q$ (cf. Eq.~\eqref{Wigner222BB}) and $-i(\hbar/2)\, \partial_p$ (cf. Eq.~\eqref{Wigner222CC}), respectively.
An equivalent Wigner continuity equation is cast in the form of Eq.~\eqref{alexquaz51}, \begin{equation} \label{helps}
\partial_t W= \sum_{\eta=0}^{\infty}\frac{(-1)^{\eta}\hbar^{2\eta}}{2^{2\eta}(2\eta+1)!} \, \left\{
\left[\partial_q^{2\eta+1}V(q)\right]\,\partial_p^{2\eta+1}W
-
\left[\partial_p^{2\eta+1}K(p)\right]\,\partial_q^{2\eta+1}W
\right\},\end{equation}
from which one has
\begin{equation}
J_q(q,\,p;\,t) = +\sum_{\eta=0}^{\infty} \left(\frac{i\,\hbar}{2}\right)^{2\eta}\frac{1}{(2\eta+1)!} \, \left[\partial_p^{2\eta+1} K(p)\right]\,\partial_q^{2\eta}W(q,\,p;\,t),
\label{alexquaz500BB}
\end{equation}
and
\begin{equation}
J_p(q,\,p;\,t) = -\sum_{\eta=0}^{\infty} \left(\frac{i\,\hbar}{2}\right)^{2\eta}\frac{1}{(2\eta+1)!} \, \left[\partial_q^{2\eta+1} V(q)\right]\,\partial_p^{2\eta}W(q,\,p;\,t),\label{alexquaz500CC}
\end{equation}
where, again, the role of the Planck constant, $\hbar$, is evinced in driving the quantum contributions coupled to the Hamiltonian non-linear dynamics.

Given the above clarification, $\hbar$ can now be absorbed by dimensionless variables, $x$ and $k$, related with the physical ones, $q$ and $p$, by $x = \left(m\,\omega\,\hbar^{-1}\right)^{1/2} q$ and $k = \left(m\,\omega\,\hbar\right)^{-1/2}p$, respectively. Hence the phase-space dynamics could be depicted by a Hamiltonian system simplified into the form of 
\begin{equation}\label{sm}
\mathcal{H}(x,\,k) = \mathcal{K}(k) + \mathcal{V}(x),
\end{equation}
such that $\mathcal{H} = (\hbar \omega)^{-1} H$, $\mathcal{V}(x) = (\hbar \omega)^{-1} V\left(\left(m\,\omega\,\hbar^{-1}\right)^{-1/2}x\right)$ and $\mathcal{K}(k) = (\hbar \omega)^{-1} K\left(\left(m\,\omega\,\hbar\right)^{1/2}k\right)$.
In this case, $m$ and $\omega$ are auxiliary mass and angular frequency parameters, and
the Wigner function can also be cast into the dimensionless form of $\mathcal{W}(x, \, k;\,\omega t) \equiv \hbar\, W(q,\,p;\,t)$\footnote{Such that, 
\begin{equation}
\int^{+\infty}_{-\infty} \hspace{-.4 cm}{dx}\int^{+\infty}_{-\infty} \hspace{-.4 cm}{dk} \,\mathcal{W}(x, \, k;\,\omega t) =
\int^{+\infty}_{-\infty} \hspace{-.4 cm}{dq}\,\int^{+\infty}_{-\infty} \hspace{-.4 cm}{dp} \,W(q,\,p;\,t)
=1,\end{equation}
and wave functions, $\phi(x,\,\tau)$, are consistently normalized by
\begin{equation}
\int^{+\infty}_{-\infty} \hspace{-.4 cm}{dx}\,\vert\phi(x;\,\tau)\vert^2 =\int^{+\infty}_{-\infty}\hspace{-.4 cm}{dq}\,\vert\psi(q;\,t)\vert^2 = 1.
\end{equation}}, i.e.
\begin{eqnarray}\label{alexDimW}
\mathcal{W}(x, \, k;\,\tau) &=& \pi^{-1} \int^{+\infty}_{-\infty} \hspace{-.35cm}dy\,\exp{\left(2\, i \, k \,y\right)}\,\phi(x - y;\,\tau)\,\phi^{\ast}(x + y;\,\tau),
\end{eqnarray}
with $y = \left(m\,\omega\,\hbar^{-1}\right)^{1/2} s$, and a dimensionless time variable, $\tau = \omega t$.
The associated (also dimensionless) Wigner flow contributions, now in the form of $\mbox{\boldmath $\mathcal{J}$} =\mathcal{J}_x\,\hat{x} + \mathcal{J}_k\,\hat{k}$, reproduce a flow field connected to the Wigner function dynamics through the
Wigner current derivatives 
\begin{eqnarray}
\label{imWA21}\partial_x\mathcal{J}_x(x, \, k) &=& +\sum_{\eta=0}^{\infty} \left(\frac{i}{2}\right)^{2\eta}\frac{1}{(2\eta+1)!} \, \left[\partial_k^{2\eta+1}\mathcal{K}(k)\right]\,\partial_x^{2\eta+1}\mathcal{W}(x, \, k),
\\
\label{imWB21}\partial_k\mathcal{J}_k(x, \, k) &=& -\sum_{\eta=0}^{\infty} \left(\frac{i}{2}\right)^{2\eta}\frac{1}{(2\eta+1)!} \, \left[\partial_x^{2\eta+1}\mathcal{V}(x)\right]\,\partial_k^{2\eta+1}\mathcal{W}(x, \, k),
\end{eqnarray}
where the time-dependence was left implicit to phase-space coordinates $x\equiv x(\tau)$ and $k\equiv k(\tau)$ \cite{Case}.
Results as described by Eqs.~\eqref{imWA21} and \eqref{imWB21} encompass all the contributions due to quantum corrections of order $\mathcal{O}(\hbar^{2\eta})$ and, as from Eq.~\eqref{alexquaz51},
define the stationarity quantifier, 
\begin{equation}\label{altz51dim}
\mbox{\boldmath $\nabla$}_{\xi}\cdot\mbox{\boldmath $\mathcal{J}$} = - {\partial_{\tau} \mathcal{W}},
\end{equation}
which, in terms of Eqs.~\eqref{imWA21} and \eqref{imWB21}, and with $\mbox{\boldmath $\xi$}= \xi_x \hat{x} + \xi_k \hat{k} = \{x,\,k\}$, is written as
\begin{equation} \label{althelps}
\mbox{\boldmath $\nabla$}_{\xi}\cdot\mbox{\boldmath $\mathcal{J}$} = \sum_{\eta=0}^{\infty}\frac{(-1)^{\eta}}{2^{2\eta}(2\eta+1)!} \, \left\{
\left[\partial_x^{2\eta+1}\mathcal{V}(x)\right]\,\partial_k^{2\eta+1}\mathcal{W}
-
\left[\partial_k^{2\eta+1}\mathcal{K}(k)\right]\,\partial_x^{2\eta+1}\mathcal{W}
\right\},\end{equation}
whose vanishing means stationarity. Given that the classical regime \cite{Case,Ballentine} limit is identified from the $\eta = 0$ contribution, such that
\begin{equation}
\mathcal{J}^{\mathcal{C}}_x(x,\, k)= +\mathcal{W}(x,\, k)\, \partial_k \mathcal{K}(k)\,\label{alexquaz500BB2}
\end{equation}
and
\begin{equation}
\mathcal{J}^{\mathcal{C}}_k(x,\, k) = -\,\mathcal{W}(x,\, k)\,\partial_x \mathcal{V}(x),
\label{alexquaz500CC2}
\end{equation}
one has
\begin{equation} \label{althelps}
\mbox{\boldmath $\nabla$}_{\xi}\cdot\mbox{\boldmath $\mathcal{J}$}^{\mathcal{C}} = 
\left[\partial_x\mathcal{V}(x)\right]\,\partial_k\mathcal{W} 
-
\left[\partial_k\mathcal{K}(k)\right]\,\partial_x\mathcal{W} =
\left[\partial_x\mathcal{H}(x,\,k)\right]\,\partial_k\mathcal{W} 
-
\left[\partial_k\mathcal{H}(x,\,k)\right]\,\partial_x\mathcal{W} ,\end{equation}
which vanishes, for instance, for the Wigner function identified by $\mathcal{W} \equiv \mathcal{W}(\mathcal{H}(x,\, k))$ (with $\mathcal{H}(x,\, k)$ as from Eq.~\eqref{sm}), as it happens for phases-space classical thermodynamic ensembles (originally addressed by the seminal Wigner's paper \cite{Wigner}), with $\mathcal{W}(x,\, k)\propto\exp[-\mathcal{H}(x,\, k)(\hbar\omega/k_B T)]$, where $k_B$ is the Boltzmann constant and $T$ is the ensemble temperature.
Specifically for these cases, with $\mathcal{W} \equiv \mathcal{W}(\mathcal{H}(x,\, k))$, with quantum corrections from $\eta \geq 1$ contributions modifying the above pattern through Eq.~\eqref{althelps}, the non-stationarity is a typical quantum effect. 
Of course, this is not true for generic Wigner distributions, where non-stationarity results from both classical ($\mbox{\boldmath $\nabla$}_{\xi}\cdot\mbox{\boldmath $\mathcal{J}$}^{\mathcal{C}}\neq 0$) and quantum ($\mbox{\boldmath $\nabla$}_{\xi}\cdot(\mbox{\boldmath $\mathcal{J}$}-\mbox{\boldmath $\mathcal{J}$}^{\mathcal{C}})\neq 0$) contributions.

Otherwise, the quantum back reaction driven by $\eta \geq 1$ contributions (as from Eqs.~\eqref{imWA21} and \eqref{imWB21}) can be more clearly evinced by a Liouvillianity vector field divergence operator.
In fact, by admitting that a constraint between $\mbox{\boldmath $\mathcal{J}$}$ and $\mathcal{W}$ can be set in terms of $\mathbf{w} = \mbox{\boldmath $\mathcal{J}$}/\mathcal{W}$, i.e. a kind of {\em quantum analog} of the classical velocity, $\mathbf{v}_{\xi(\mathcal{C})}$, the non-Liouvillianity, identified by $\mbox{\boldmath $\nabla$}_{\xi} \cdot \mathbf{w} \neq 0$, is a typical quantum effect quantified by\footnote{Given that $\mbox{\boldmath $\nabla$}_{\xi}\cdot\mbox{\boldmath $\mathcal{J}$} = \mathcal{W}\,\mbox{\boldmath $\nabla$}_{\xi}\cdot\mathbf{w}+ \mathbf{w}\cdot \mbox{\boldmath $\nabla$}_{\xi}\mathcal{W}$ \cite{Steuernagel3}, $\mbox{\boldmath $\nabla$}_{\xi}\cdot\mbox{\boldmath $\mathcal{J}$}$ and $\mbox{\boldmath $\nabla$}_{\xi} \cdot \mathbf{w}$ are mutually connected by the relation
\begin{equation}
\mbox{\boldmath $\nabla$}_{\xi} \cdot \mathbf{w} = \frac{\mathcal{W}\, \mbox{\boldmath $\nabla$}_{\xi}\cdot \mbox{\boldmath $\mathcal{J}$}- \mbox{\boldmath $\mathcal{J}$}\cdot\mbox{\boldmath $\nabla$}_{\xi}{\mathcal{W}}}{\mathcal{W}^2}.
\label{zeqnz59}
\end{equation}}.
\begin{equation}\label{altdiv2}
\mbox{\boldmath $\nabla$}_{\xi} \cdot \mathbf{w} = \sum_{\eta=1}^{\infty}\frac{(-1)^{\eta}}{2^{2\eta}(2\eta+1)!}
\left\{
\left[\partial_k^{2\eta+1}\mathcal{K}(k)\right]\,
\partial_x\left[\frac{1}{\mathcal{W}}\partial_x^{2\eta}\mathcal{W}\right]
-
\left[\partial_x^{2\eta+1}\mathcal{V}(x)\right]\,
\partial_k\left[\frac{1}{\mathcal{W}}\partial_k^{2\eta}\mathcal{W}\right]
\right\}, ~~~\end{equation}
whose vanishing means Liouvillianity\footnote{Here one notices that the $\eta = 0$ contribution is indeed irrelevant since the classical limit, $\mbox{\boldmath $\nabla$}_{\xi} \cdot \mathbf{v}_{\xi(\mathcal{C})} = 0$, follows from
$$\frac{(-1)^{\eta}}{2^{2\eta}(2\eta+1)!}
\left\{
\left[\partial_k^{2\eta+1}\mathcal{K}(k)\right]\,
\partial_x\left[\frac{1}{\mathcal{W}}\partial_x^{2\eta}\mathcal{W}\right]
-
\left[\partial_x^{2\eta+1}\mathcal{V}(x)\right]\,
\partial_k\left[\frac{1}{\mathcal{W}}\partial_k^{2\eta}\mathcal{W}\right]
\right\}\bigg{\vert}_{\eta=0}=0.$$}.

\section{Gaussian distribution ensembles}

Wigner functions with $x$ and $k$ contributions decoupled into the form of $\mathcal{W}(x,\,k) \equiv \vert\phi(x;\,\tau)\vert^2 \vert\psi(k;\,t)\vert^2$ lead to substantial simplifications on the above results. This can be verified for Wigner functions described by Gaussian distributions.
In fact, for a normalized Gaussian Wigner function cast into a dimensionless form given by
\begin{equation}\label{gau}
\mathcal{G}_\alpha(x,\,k) = \frac{\alpha^2}{\pi}\, \exp\left[-\alpha^2\left(x^2+ k^2\right)\right],
\end{equation}
for the Hamiltonian in the form like Eq.~(\ref{sm}), the following associated Wigner flow contributions are obtained \cite{Novo21A},
\begin{eqnarray}
\label{imWA2}\partial_x\mathcal{J}_x(x, \, k) &=& +\sum_{\eta=0}^{\infty} \left(\frac{i}{2}\right)^{2\eta}\frac{1}{(2\eta+1)!} \, \left[\partial_k^{2\eta+1}\mathcal{K}(k)\right]\,\partial_x^{2\eta+1}\mathcal{G}_{\alpha}(x, \, k),
\\
\label{imWB2}\partial_k\mathcal{J}_k(x, \, k) &=& -\sum_{\eta=0}^{\infty} \left(\frac{i}{2}\right)^{2\eta}\frac{1}{(2\eta+1)!} \, \left[\partial_x^{2\eta+1}\mathcal{V}(x)\right]\,\partial_k^{2\eta+1}\mathcal{G}_{\alpha}(x, \, k).
\end{eqnarray}
From Gaussian relations with Hermite polynomials of order $n$, $\mbox{\sc{h}}_n$, one has \cite{Novo21A}
\begin{equation}
\partial_\zeta^{2\eta+1}\mathcal{G}_{\alpha}(x, \, k) = (-1)^{2\eta+1}\alpha^{2\eta+1}\,\mbox{\sc{h}}_{2\eta+1} (\alpha \zeta)\, \mathcal{G}_{\alpha}(x, \, k),
\end{equation}
for $\zeta = x,\, k$, which can be reintroduced in Eqs.~\eqref{imWA2} and \eqref{imWB2} to return convergent series expansions. These allow for recasting the Wigner flow expressions into an analytical form, which accounts for the overall quantum modifications, i.e. for $\eta$ from $1$ to $\infty$ into Eqs.~(\ref{imWA2})-(\ref{imWB2}).
In particular, for the quantum systems where $\mathcal{V}$ and $\mathcal{K}$ derivatives result in
\begin{eqnarray}
\label{t111}
\partial_x^{2\eta+1}\mathcal{V}(x) &=& \lambda^{2\eta+1}_{(x)} \, \upsilon(x),\\
\label{t222}
\partial_k^{2\eta+1}\mathcal{K}(k) &=& \mu^{2\eta+1}_{(k)} \, \kappa(k),
\end{eqnarray}
where $\lambda$, $\upsilon$, $\mu$, and $\kappa$ are arbitrary auxiliary functions, it can be straightforwardly verified that, once they are substituted into Eqs.~\eqref{imWA2} and \eqref{imWB2}, the above expressions lead to
\begin{eqnarray}
\label{imWA3}\partial_x\mathcal{J}_x(x, \, k) &=& (+2i) \kappa(k)\,\mathcal{G}_{\alpha}(x, \, k)\,\sum_{\eta=0}^{\infty} \left(\frac{i\,\alpha\,\mu_{(k)}}{2}\right)^{2\eta+1}\frac{1}{(2\eta+1)!} \, \mbox{\sc{h}}_{2\eta+1} (\alpha x),\\
\label{imWB3}\partial_k\mathcal{J}_k(x, \, k) &=& (-2i) \upsilon(x)\,\mathcal{G}_{\alpha}(x, \, k)\sum_{\eta=0}^{\infty} \left(\frac{i\,\alpha\, \lambda_{(x)}}{2}\right)^{2\eta+1}\frac{1}{(2\eta+1)!} \, \mbox{\sc{h}}_{2\eta+1} (\alpha k).\end{eqnarray}
Finally, by noticing that \cite{Novo21A}
\begin{equation}
\sum_{\eta=0}^{\infty}\mbox{\sc{h}}_{2\eta+1} (\alpha \zeta)\frac{s^{2\eta+1}}{(2\eta+1)!} = \sinh(2s\,\alpha\zeta) \exp[-s^2],
\end{equation}
one gets \cite{Novo21A}
\begin{eqnarray}
\label{imWA4}\partial_x\mathcal{J}_x(x, \, k) &=& -2 \kappa(k)\,\sin\left(\alpha^2 \mu_{(k)}\,x\right)\,\exp[+\alpha^2 \mu^2_{(k)}/4]\,\mathcal{G}_{\alpha}(x, \, k)\,
,\\
\label{imWB4}\partial_k\mathcal{J}_k(x, \, k) &=& +2 \upsilon(x)\,\sin\left(\alpha^2 \lambda_{(x)}\,k\right)\,\exp[+\alpha^2 \lambda^2_{(k)}/4]\,\mathcal{G}_{\alpha}(x, \, k),
\end{eqnarray}
which, as expected from the convergence of the above presented series expansions, give rise to an analytical form for the stationarity quantifier, $\mbox{\boldmath $\nabla$}_{\xi}\cdot \mbox{\boldmath $\mathcal{J}$}$, which can be manipulated to give the Liouvillian quantifier, $\mbox{\boldmath $\nabla$}_{\xi} \cdot \mathbf{w}$.

It is worth mentioning that the convergent expressions (cf. Eqs.~\eqref{imWA4} and \eqref{imWB4}) obtained from the infinite series expansions for the Wigner currents (cf. Eqs.~\eqref{imWA2} and \eqref{imWB2}), couple quantum fluctuations to non-linear effects, without any truncating stratagem. Therefore, through the Wigner framework, here evaluated for Gaussian distributions, QM cannot be decoupled from non-linearity. In the next section, the same conclusion will extend to gamma and Laplacian distribution ensembles.

\section{Gamma and Laplacian distribution ensembles}

In probability theory and statistics, the gamma distribution is a two-parameter family of continuous probability distributions. Exponential distributions, Erlang distributions, and chi-squared distributions are special cases of the gamma distribution, which assumes a normalized phase-space dimensionless form given by
\begin{eqnarray}\label{gaa}
\mbox{G}^{(a,b)}_{(\alpha,\beta)}(x,\,k) &=&\frac{x^{a-1}k^{b-1}}{\Gamma(a)\Gamma(b)}\beta^{b}\alpha^{a} \exp(-\alpha x - \beta k) \nonumber\\&=&(-1)^{a+b}\frac{\beta^{b}\alpha^{a} }{\Gamma(a)\Gamma(b)} \partial_\alpha^{a-1}\partial_\beta^{b-1}\exp(-\alpha x - \beta k),
\end{eqnarray}
where $\Gamma(\dots)$ is the gamma function.

Gamma distributions in one-dimension are used to model a wide variety of phenomena composed by several sub-events
which occurs in sequence, with step-time driven by the exponential distribution (with rate $\alpha$ or $\beta$ in Eq.~\eqref{gaa}). This includes, for instance, the waiting time of cell-division events \cite{26} or the number of compensatory mutations for a given mutation \cite{27}.

For the Hamiltonian in the form of Eq.~\eqref{sm}, with the simplifications from Eqs.~\eqref{t111} and \eqref{t222},
the following associated Wigner flow contributions are obtained as
\begin{eqnarray}
\label{imWA2m}\partial_x\mathcal{J}_x(x, \, k) &=& +\sum_{\eta=0}^{\infty} \left(\frac{i}{2}\right)^{2\eta}\frac{1}{(2\eta+1)!} \left[\partial_k^{2\eta+1}\mathcal{K}(k)\right]\,\partial_x^{2\eta+1}\mbox{G}^{(a,b)}_{(\alpha,\beta)}(x,\,k)\\
&=& +(-1)^{a+b}\kappa(k)\,\frac{2\beta^{b}\alpha^{a}}{i\Gamma(a)\Gamma(b)}\partial_\alpha^{a-1}\partial_\beta^{b-1} \left[\sum_{\eta=0}^{\infty}\frac{(-i\alpha\mu_{(k)}/2)^{2\eta+1}}{(2\eta+1)!} \exp(-\alpha x - \beta k)\right], \nonumber\\
\label{imWB2m}\partial_k\mathcal{J}_k(x, \, k) &=& -\sum_{\eta=0}^{\infty} \frac{1}{(2\eta+1)!} \left(\frac{i}{2}\right)^{2\eta} \left[\partial_x^{2\eta+1}\mathcal{V}(x)\right]\,\partial_k^{2\eta+1}\mbox{G}^{(a,b)}_{(\alpha,\beta)}(x,\,k)\\&=& -(-1)^{a+b}\upsilon(x)\,\frac{2\beta^{b}\alpha^{a} }{i\Gamma(a)\Gamma(b)}\partial_\alpha^{a-1}\partial_\beta^{b-1} \left[\sum_{\eta=0}^{\infty}\frac{(-i\beta \lambda_{(x)}/2)^{2\eta+1}}{(2\eta+1)!} \, \exp(-\alpha x - \beta k)\right], \nonumber
\end{eqnarray}
which can be cast in the simple form of
\begin{eqnarray}
\label{imWA2n}\partial_x\mathcal{J}_x(x, \, k) &=& +2\kappa(k)\,(-1)^{a}k^{b-1}\frac{\beta^{b}\alpha^{a}}{\Gamma(a)\Gamma(b)}\partial_\alpha^{a-1}\left[\sin\left(\frac{\alpha\mu_{(k)}}{2}\right)\exp(-\alpha x - \beta k)\right], \\
\label{imWB2n}\partial_k\mathcal{J}_k(x, \, k) &=& -2\upsilon(x)\,(-1)^{b}\,x^{a-1}\frac{\beta^{b}\alpha^{a} }{\Gamma(a)\Gamma(b)}\partial_\beta^{b-1} 
\left[\sin\left(\frac{\beta \lambda_{(x)}}{2}\right) \, \exp(-\alpha x - \beta k)\right]. 
\end{eqnarray}

As supposed, depending on the explicit form of the Hamiltonian, these flow equations can be manipulated to give the Liouvillianity quantifier, $\mbox{\boldmath $\nabla$}_{\xi} \cdot \mathbf{w}$ and the complete pattern of the associated Wigner flow.

However, in contrast to Gaussian distribution ensembles, for which the phase-space coordinate support is given by $x,\, k \in \{-\infty,\,+\infty\}$, for gamma distribution ensembles, the corresponding support is given by $x,\, k \in \{0,\,+\infty\}$.
Otherwise, depending on the convenience, gamma distributions can also be replaced by Laplacian distributions, written as $(1/4)\mbox{G}^{(a,b)}_{(\alpha,\beta)}(\vert x\vert,\,\vert k\vert)$. In this case, with the support extended to $x,\, k \in \{-\infty,\,+\infty\}$, one would have
\begin{eqnarray}
\label{imWA2bn}\partial_x\mathcal{J}_x(x, \, k) &=& +\frac{\kappa(k)}{2}\,(-1)^{a}\vert k\vert ^{b-1}\frac{\beta^{b}\alpha^{a}}{\Gamma(a)\Gamma(b)}\partial_\alpha^{a-1}\left[\sin\left(\frac{\alpha\mu_{(k)}}{2}\right)\exp(-\alpha \vert x\vert - \beta \vert k\vert )\right], \\
\label{imWB2bn}\partial_k\mathcal{J}_k(x, \, k) &=& -\frac{\upsilon(x)}{2}\,(-1)^{b}\,\vert x\vert ^{a-1}\frac{\beta^{b}\alpha^{a} }{\Gamma(a)\Gamma(b)}\partial_\beta^{b-1} 
\left[\sin\left(\frac{\beta \lambda_{(x)}}{2}\right) \, \exp(-\alpha \vert x\vert - \beta \vert k\vert)\right],\quad 
\end{eqnarray}
which, in correspondence with Eqs.~\eqref{imWA2n} and \eqref{imWB2n}, can be relevant in the investigation of statistical implications in driving quantum fluctuations.

\section{Examples of Hamiltonian systems: typical and modified prey-predator dynamics}

The LV dynamical equations for prey-predator systems \cite{LV1,LV2}, once related to a Hamiltonian dynamics, are suitable for showing how quantum mechanical modifications appear. In spite of being usually considered for describing the behavior of macroscopic ecosystems \cite{PRE-LV,SciRep02,PRE-LV2,RPSA-LV,Anna}, the LV system has also been discussed in a wide range of microscopic scenarios which include, for instance, the description of stability criteria for microbiological communities \cite{Nature01, Nature02}, the emergence of phase transitions in finite microscopic systems \cite {PRE-LV3}, and the support for the dynamically driven stochastic systems \cite{Allen,Grasman}.
The phenomenology of prey-predator dynamics of the 2-dim LV system can be found once it is put into the form of
\begin{eqnarray}\label{Ham2bbbA}
d{y}/d\tau &=& (y\,z - y),\\
\label{Ham2bbbB}
d{z}/d\tau &=& g\,(z - y\,z),
\end{eqnarray}
where $g > 0$ is an arbitrary anisotropy parameter for $y(\tau)$ and $z(\tau)$ describing predator and prey populations, respectively. In this case, the populations are normalized by their corresponding time-averaged mean populations which set equilibrium points at $y = z = 1$. 

\subsection{Typical LV Hamiltonian formulation -- Classical features}

Eqs.~\eqref{Ham2bbbA}-\eqref{Ham2bbbB} have a correspondence with Hamiltonian pattern given by
\begin{equation}\label{altHam}
\mathcal{H}(x,\,k) = g \,x + k + g\, e^{-x} + e^{-k},
\end{equation}
for which the condition $\partial^2 \mathcal{H} / \partial x \, \partial k = 0$ is satisfied.
In this case, Hamilton's equations result in
\begin{eqnarray}
\label{altHam2B}
d{x}/d\tau &=& \{x,\mathcal{H}\}_{PB} = 1-e^{-k},\\
\label{altHam2C}
d{k}/d\tau &=& \{k,\mathcal{H}\}_{PB} = g\,e^{-x} - g,
\end{eqnarray}
which drive $x$ and $k$ oscillations correlated to the number of prey and predator species, $y$ and $z$, by $y = e^{-x}$ and $z = e^{-k}$. Solutions correspond to ecological coexistence chains theoretically depicted by phase-space closed orbits,
$\mathcal{H}(x,\,k) = \epsilon$, with $\epsilon \in (g+1,\infty)$, which correspond to $y-z$ phase-space implicit level curves given by
\begin{eqnarray}\label{Ham2ccc}
\epsilon = g\,y + z - \ln(y^g\,z),
\end{eqnarray}
as they follow from straightforward integration of
\begin{eqnarray}
-g \,\frac{d{y}}{d z} &=& \frac{z - 1}{z}\frac{y}{y-1}.
\label{Hs}
\end{eqnarray}

Semi-analytical solutions for Eqs.~\eqref{Ham2bbbA}-\eqref{Ham2bbbB} are only admitted for the isotropic coordinate version of the system (i.e. with $g = 1$), with the phase-space trajectories corresponding to the parametric curve,
\begin{eqnarray}\label{Ham2B}
 y(\mathcal{T})&=&\frac{\mathcal{T} \pm \sqrt{\mathcal{T}^2-4 e^{\mathcal{T} - \epsilon}}}{2},\\
 z(\mathcal{T})&=&\frac{\mathcal{T} \mp \sqrt{\mathcal{T}^2-4 e^{\mathcal{T} - \epsilon}}}{2},
\end{eqnarray}
with the dynamical constraint,
\begin{eqnarray}\label{Ham2C}
\dot{\mathcal{T}}^2 - {\mathcal{T}^2-4 e^{\mathcal{T} - \epsilon}}=0,
\end{eqnarray}
where ``{\em dots}'' correspond to ${\tau}$ derivatives, $d/d{\tau}$, and $\epsilon$ is a (dimensionless) constant of motion.

For $\Gamma_{\epsilon}$ closed paths, the periodic ciclic dynamics is identified by the coordinate integrals evaluated as
\begin{eqnarray}\label{Ham2ddd}
\int_0^{T} y(\tau)\,d\tau = \int_0^{T} z(\tau)\,d\tau = \int_0^{T} y(\tau)\,z(\tau)\,d\tau=T, \end{eqnarray}
which is obtained from a direct $\tau$ integration of Eqs.~\eqref{Ham2bbbA}-\eqref{Ham2bbbB} along the period of time $T$ and by straightforwardly noticing that
\begin{eqnarray}\label{auzzz}
0 &=& \oint_{\Gamma_{{\epsilon}}} dF(y) = \oint_{\Gamma_{{\epsilon}}} dy\,y^{-1} = \int_0^{T}(dy/d\tau) y^{-1}\,d\tau =\int_0^{T} (z(\tau)-1)\,d\tau\nonumber\\
&=&\oint_{\Gamma_{{\epsilon}}} dG(z) = \oint_{\Gamma_{{\epsilon}}} dz\,z^{-1} = \int_0^{T}(dz/d\tau) z^{-1}\,d\tau =\int_0^{T} (y(\tau)-1)\,d\tau, \end{eqnarray}
for arbitrary $F(y)$ and $G(z)$. Additionally, from Green's theorem, one has the $y-z$ plane area enclosed by $\Gamma_{{\epsilon}}$, $\mathcal{A}_{\Gamma_{\epsilon}}$, given either by
\begin{eqnarray}
\mathcal{A}_{\Gamma_{\epsilon}}&=& -\oint_{\Gamma_{{\epsilon}}} y\,dz = g\, \int_0^{T}y(\tau) \,z(\tau)(y(\tau)-1)\,d\tau = g\, \int_0^{T}y(\tau) \,(y(\tau)-1)\,d\tau\nonumber\\
&=&g\,\int_0^{T} \,(y(\tau)-1)^2\,d\tau, \end{eqnarray}
or by
\begin{eqnarray}
\mathcal{A}_{\Gamma_{\epsilon}} &=& + \oint_{\Gamma_{{\epsilon}}} z\,dy = \,\int_0^{T}y(\tau) \,z(\tau)(z(\tau)-1)\,d\tau= \,\int_0^{T} \,z(\tau)(z(\tau)-1)\,d\tau\nonumber\\
&=&\, \int_0^{T} \,(z(\tau)-1)^2\,d\tau,\end{eqnarray}
where manipulations involving Eqs.~\eqref{Ham2bbbA}-\eqref{Ham2bbbB} and \eqref{Ham2ddd}-\eqref{auzzz} have been used.

Once again from Green's theorem, one also has the $x-k$ plane area enclosed by $\Gamma_{H}$, ${A}_{\Gamma_{H}}$, given by
\begin{eqnarray}\label{Quant}
{A}_{\Gamma_{H}}=\oint_{\Gamma_{{H}}} dx\, k &=& - \oint_{\Gamma_{{H}}} dk\, x = 
\frac{1}{2}\left[\oint_{\Gamma_{{H}}} dx\, k- \oint_{\Gamma_{{H}}} dk\, x\right]\nonumber\\
&=& \frac{1}{2}\int_0^{T} \,\left\{(z(\tau)-1)\ln[z(\tau)]+g (y(\tau)-1)\ln[y(\tau)]\right\}\,d\tau,
\end{eqnarray}
and noticing from Eqs.~\eqref{Ham2bbbA}-\eqref{Ham2bbbB} that
\begin{eqnarray}
g \int_0^{T} \,(z(\tau)-1)\ln[y(\tau)]d\tau = \int_0^{T} \,(y(\tau)-1)\ln[z(\tau)]d\tau = 0,
\end{eqnarray}
for $\mathcal{H}(x,\,k) = \epsilon$, one can write,
\begin{eqnarray}
\oint_{\Gamma_{{H}=\epsilon}} dx\, k &=& \int_0^{T} \,\left\{(z(\tau)-1)+ (y(\tau)-1)\right\}\ln[z(\tau)\,y^g(\tau)]\,d\tau\nonumber\\
&=& \frac{1}{2}\int_0^{T} \,\left\{(z(\tau)-1)+ (y(\tau)-1)\right\}\left\{\left[g\,y(\tau)+z(\tau)\right]-\epsilon\right\}\,d\tau\nonumber\\
&=& \dots\nonumber\\
&=& \mathcal{A}_{\Gamma_{\epsilon}},
\end{eqnarray}
providing the constraint $\mathcal{A}_{\Gamma_{\epsilon}} = {A}_{\Gamma_{H=\epsilon}}$ between $y-z$ and $x-k$ enclosed areas.

Besides sedimenting the phase-space quantum analysis that follows, the above parameterization, once supported by a Hamiltonian dynamics and by results from Eq.~\eqref{Ham2ddd}, suggests an equivalent Bohr-Sommerfeld quantization rule as
\begin{equation}\label{EqHam2B6}
2\pi\ell =\oint_{\Gamma_{{H}=\epsilon}} dx\, k = {A}_{\Gamma_{H=\epsilon}}=\mathcal{A}_{\Gamma_{\epsilon}},
\end{equation}
where $\ell$ is an integer quantum number, $\ell = 1,\,2,\, \dots$ and the reduced Planck constant was set as $\hbar = 1$.
More relevant, however, is the connection which can be established with the energy-like parameter $\epsilon$, which can also be quantized. In fact, from Eq.~\eqref{Quant} one has
\begin{eqnarray}\label{Quant2}
\frac{d{A}}{d\tau} &=& 
\frac{1}{2}\left[k \frac{dx}{d\tau}\,- x \frac{dk}{d\tau}\right]= \frac{\theta}{2}\left[k (1-e^{-\theta k})+ g\, x (1-e^{-\theta x})\right]\nonumber\\
&=& \frac{\theta}{2}\frac{d~}{d\theta}\left[\theta(g \,x + k) + g\, e^{-\theta x} + e^{-\theta k}\right]\bigg{\vert}_{\theta=1}
\nonumber\\
&=& \frac{\theta}{2}\frac{d~}{d\theta}\epsilon(\theta)\bigg{\vert}_{\theta=1},
\end{eqnarray}
which once integrated along $\Gamma_{H= \epsilon}$ closed paths, leads to
\begin{equation}\label{Quant3}
\mathcal{A}_{\Gamma_{\epsilon}} = {A}_{\Gamma_{H=\epsilon}} = T\,\frac{\theta}{2}\frac{d~}{d\theta}\epsilon(\theta)\bigg{\vert}_{\theta=1} = 2\pi\ell,
\end{equation}
which sets the Bohr-Sommerfeld quantized trajectories for $\mathcal{H}(x,\,k)$\footnote{
In fact, for any Hamiltonian $\mathcal{H}(x,\,k)$ set as $\mathcal{H}(x,\,k;\,\theta) \equiv \mathcal{H}(\theta x,\,\theta k) \equiv \epsilon(\theta)$ one has
\begin{eqnarray}\label{Quant4}
\frac{d{A}}{d\tau} &=& 
\frac{1}{2}\left[k \frac{dx}{d\tau}\,- x \frac{dk}{d\tau}\right]= \frac{1}{2}\left[k \frac{\partial}{\partial k}+x \frac{\partial}{\partial x} \right]\mathcal{H}(\theta x,\,\theta k)\nonumber\\
&=& \frac{1}{2}\mbox{\boldmath $\xi$}\cdot\mbox{\boldmath $\nabla$}_{\xi}\mathcal{H}(\theta x,\,\theta k)
= \frac{\xi}{2}\frac{d~}{d\xi}\mathcal{H}(\theta x,\,\theta k) = \frac{\xi}{2}\frac{d~}{d\xi}\mathcal{H}(\theta \xi \cos(\vartheta),\,\theta \xi \sin(\vartheta))\nonumber\\
&=& \frac{\theta}{2}\frac{d~}{d\theta}\mathcal{H}(\theta \xi \cos(\vartheta),\theta \xi \sin(\vartheta))= \frac{\theta}{2}\frac{d~}{d\theta}\mathcal{H}(\theta x,\,\theta k)\nonumber\\
&=& \frac{\theta}{2}\frac{d~}{d\theta}\epsilon(\theta).
\end{eqnarray}
where the Hamiltonian invariance with respect to the permutation $\xi \leftrightarrow \theta$ have been considered in the third step above.}.

\subsection{Modified LV Hamiltonian formulation -- Classical features}

An extension of the typical LV description \cite{LV1,LV2,Novo21B,Novo21B2,Novo222} can also be evaluated from the Hamiltonian written in the dimensionless form of
\begin{equation}\label{Original}
\mathcal{H}_{\mbox{\tiny M}}(x,\,k) = \cosh(k) + g\,\cosh(x),
\end{equation}
which exhibits such a dynamical structure reflected by classical equations of motion in the form of
\begin{eqnarray}\label{Ham2Bs}
d{x}/d\tau &=& \{x,\mathcal{H}_{\mbox{\tiny M}}\}_{PB} = \sinh(k),\\
\label{Ham2Bt}d{k}/d\tau &=& \{k,\mathcal{H}_{\mbox{\tiny M}}\}_{PB} = -g\,\sinh(x).
\end{eqnarray}
The prey-predator-like map, $y = e^{-x}$ and $z = e^{-k}$, is thus straightforwardly set up by the constraint 
\begin{eqnarray}\label{Ham2cccB}
\epsilon = \frac{1}{2}\left(g\,y + \frac{g}{y} + z + \frac{1}{z}\right),
\end{eqnarray}
with the corresponding 2-dim system obtained from Eqs.~\eqref{Ham2Bs}-\eqref{Ham2Bt} written as
\begin{eqnarray}\label{Ham2bbb22w}
d{y}/d\tau &=& \frac{1}{2}\,\left(y\,z-\frac{y}{z} \right),\\
\label{Ham2bbb22t}d{z}/d\tau &=& \frac{g}{2}\,\left(\frac{z}{y}-y\,z\right),
\end{eqnarray}
from which, for $\Gamma_{{H}_{\mbox{\tiny M}}=\epsilon}$ phase-space closed paths, one has
\begin{equation}
\int_0^{T}d\tau\, y^{-1} (d{y}/d\tau) = \int_0^{T}d\tau\, z^{-1} (d{z}/d\tau) = 0,
\end{equation}
for a periodic ciclic dynamics with period of time $T$. This leads to the coordinate integrals evaluated as
\begin{eqnarray}\label{Ham2bbb33s}
\int_0^{T}d\tau\, y &=& \int_0^{T}d\tau\, y^{-1},\\
\label{Ham2bbb33t}\int_0^{T}d\tau\, z &=& \int_0^{T}d\tau\, z^{-1},
\end{eqnarray}
and therefore, from a direct $\tau$ integration of Eq.~\eqref{Ham2cccB}, the periodic ciclic dynamics is identified by the coordinate integrals evaluated by
\begin{eqnarray}
\frac{1}{\epsilon}\int_0^{T}d\tau\, (g y+z) = \frac{1}{\epsilon}\int_0^{T}\,d\tau\, (g y^{-1}+z^{-1})=T.
\end{eqnarray}

Again, from Green's theorem, one has the $y-z$ plane area enclosed by $\Gamma_{{H}}$, $\mathcal{A}_\Gamma$, given either by
\begin{eqnarray}
\mathcal{A}_\Gamma&=& -\oint_{\Gamma_{{H}}} y\,dz = \frac{g}{2} \int_0^{T} z(\tau) \,(y^2(\tau)-1)\,d\tau,
\end{eqnarray}
or by
\begin{eqnarray}
\mathcal{A}_\Gamma &=& + \oint_{\Gamma_{{H}}} z\,dy = \frac{1}{2} \int_0^{T} y(\tau) \,(z^2(\tau)-1)\,d\tau,
\end{eqnarray}
where manipulations involving Eqs.~\eqref{Ham2bbb22w}-\eqref{Ham2bbb22t} and \eqref{Ham2bbb33s}-\eqref{Ham2bbb33t} have been used.

From a straight mathematical perspective, besides the parity symmetry exhibited by the Hamiltonian map from Eq.~\eqref{Original}, parametric solutions can be obtained for the system Eqs.~\eqref{Ham2bbb33s}-\eqref{Ham2bbb33t} as they lead to
\begin{eqnarray}\label{sols}
y(\tau) &=&\mathcal{T} \pm \sqrt{\mathcal{T}^2-\frac{\mathcal{T}}{\epsilon-\mathcal{T}}},\nonumber\\
z(\tau) &=&\mathcal{T} \mp \sqrt{\mathcal{T}^2-\frac{\mathcal{T}}{\epsilon-\mathcal{T}}},
\end{eqnarray}
with the dynamical constraint,
\begin{eqnarray}\label{Ham2CBB}
\dot{\mathcal{T}}^2 - \mathcal{T}^2(\mathcal{T} - \epsilon)^2-\mathcal{T}(\mathcal{T} - \epsilon)=0.
\end{eqnarray}

\begin{figure}[h]
\includegraphics[scale=0.8]{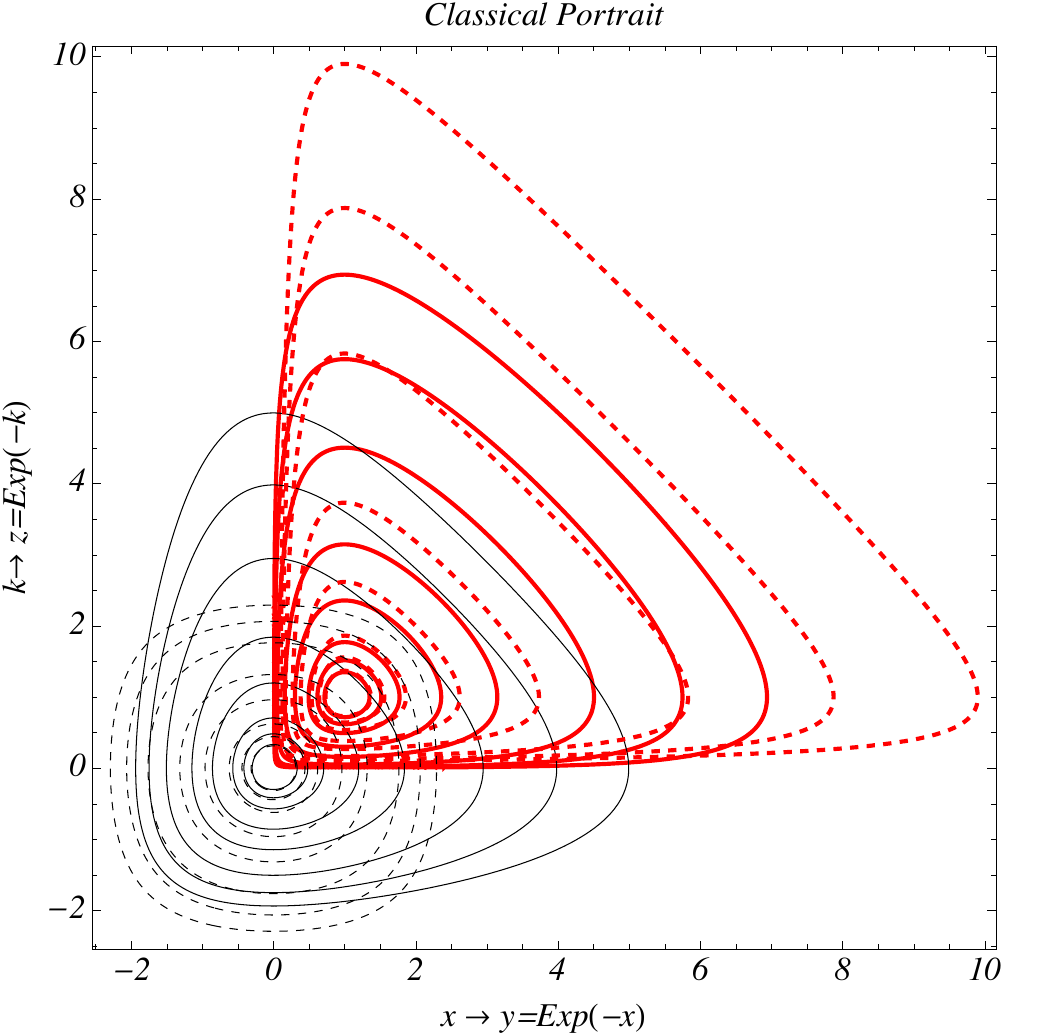}
\renewcommand{\baselinestretch}{.65}
\caption{\footnotesize
(Color online) Classical portrait of typical (solid lines) and modified (dashed lines) LV associated Hamiltonians. The phase-space $x-k$ trajectories (black thin lines) are for $\epsilon = 6,\, 5,\, 4,\, 3,\, 2.5,\, 2.2,\, 2.1$ and $2.05$, and the corresponding species distributions (red thick lines), $z$ and $y$, are identified by $y = e^{-x}$ and $z = e^{-k}$.}
\label{LVLVLV}
\end{figure}

From Fig.~\ref{LVLVLV}, one notices that, for smaller oscillation amplitudes, an approximated harmonic oscillation pattern is found in both descriptions, for which both Hamiltonians, Eqs.~\eqref{altHam} and \eqref{Original}, can be approached by
\begin{equation}\label{HamBm}
\mathcal{H}(x,\,k) = (1+g) + \frac{1}{2}\left(x^2 + k^2)\right) + \mathcal{O}(x^3,\,k^3).
\end{equation}
For the identical values of the energy parameter, $\epsilon$, the modified Hamiltonian, Eq.~\eqref{Original}, leads to increasing oscillation amplitudes with respect to the LV original formulation, which correspond to a persistence of prey and predator dominant stages.

\subsection{Wigner currents for typical LV systems -- Gaussian ensemble quantum features} 

Assuming potential implications for typical prey-predator-like dynamics, for the Hamiltonian in the form of Eq.~\eqref{altHam}, such that
\begin{eqnarray}
\label{t111B}
\partial_k^{2\eta+1}\mathcal{K}(k) &=& \delta_{\eta 0} - e^{-k},\\
\label{t222B}
\partial_x^{2\eta+1}\mathcal{V}(x) &=& g \left(\delta_{\eta 0} - e^{-x}\right), 
\end{eqnarray}
and from Eqs.~\eqref{imWA2} and \eqref{imWB2}, with manipulations that follow from Eqs.~\eqref{imWA3}-\eqref{imWB4}, the associated Wigner flow contributions are obtained for Gaussian ensembles (cf. Eq.~\eqref{gau})\footnote{With corresponding classical contributions given by
\begin{eqnarray}
\label{imWA4CC3cl}
\partial_x\mathcal{J}^{\mathcal{C}\alpha}_x(x, \, k) &=& -2 \alpha^2 \,x\left[1 - e^{-k}
\right]
\mathcal{G}_{\alpha}(x, \, k),\\
\label{imWB4CC3cl}
\partial_k\mathcal{J}^{\mathcal{C}\alpha}_k(x, \, k) &=& +2g\,\alpha^2 \,k\left[1 - e^{-x}
\right]
\mathcal{G}_{\alpha}(x, \, k).
\end{eqnarray}} \cite{Novo21B,Novo21B2},
\begin{eqnarray}
\label{imWA4CC3}
\partial_x\mathcal{J}^{\alpha}_x(x, \, k) &=& -2 \left[\alpha^2 \,x - 
\sin\left(\alpha^2\,x\right)\,e^{\frac{\alpha^2}{4}-k}
\right]
\mathcal{G}_{\alpha}(x, \, k),\\
\label{imWB4CC3}
\partial_k\mathcal{J}^{\alpha}_k(x, \, k) &=& +2g\left[\alpha^2\,k - 
\sin\left(\alpha^2\,k\right)\,e^{\frac{\alpha^2}{4}-x}
\right]
\mathcal{G}_{\alpha}(x, \, k).
\end{eqnarray}
In this case, after position or momentum integrations, one has
\begin{eqnarray}
\label{imWA4CCD4}\mathcal{J}^{\alpha}_x(x, \, k) &=& 
\mathcal{G}_{\alpha}(x, \, k)
-\frac{i\,\alpha}{2\sqrt{\pi}} \,e^{-(k+\alpha^2 k^2)}
\left\{\mbox{\sc{Erf}}\left[\alpha(x-i/2)\right]-\mbox{\sc{Erf}}\left[\alpha(x+i/2)\right]\right\},\\
\label{imWB4CCD4}\mathcal{J}^{\alpha}_k(x, \, k) &=& 
-g\,\mathcal{G}_{\alpha}(x, \, k)
+\frac{i\,g\,\alpha}{2\sqrt{\pi}} \,e^{-(x+\alpha^2 x^2)}
\left\{\mbox{\sc{Erf}}\left[\alpha(k-i/2)\right]-\mbox{\sc{Erf}}\left[\alpha(k+i/2)\right]\right\},\,\,\,\,
\end{eqnarray}
expressed in terms of error functions, $\mbox{\sc{Erf}}[\dots]$.

\subsection{Wigner currents for typical LV systems -- Gamma distribution ensemble quantum features} 

Likewise, for gamma distribution ensembles (cf. Eq.~\eqref{gaa}), from Eqs.~\eqref{imWA2m} and \eqref{imWB2m}, and manipulations which result in Eqs.~\eqref{imWA2bn}-\eqref{imWB2bn}, the following associated Wigner flow contributions are found\footnote{With corresponding classical contributions given by
\begin{eqnarray}
\label{NovoAA1cl}\partial_x\mathcal{J}_x(x, \, k) &=& +
(-1)^{a}\,\left(1-e^{-k}\right)k^{b-1}\frac{\beta^{b}\alpha^{a}}{\Gamma(a)\Gamma(b)}\partial_\alpha^{a-1}\left\{\alpha\exp(-\alpha x - \beta k)\right\}, \\
\label{NovoBB1cl}\partial_k\mathcal{J}_k(x, \, k) &=& -g\,(-1)^{b}\,\left(1-e^{-x} \right)\,x^{a-1}\frac{\beta^{b}\alpha^{a} }{\Gamma(a)\Gamma(b)}\partial_\beta^{b-1} 
\left\{\beta\, \exp(-\alpha x - \beta k)\right\}. \,\,
\end{eqnarray}
},
\begin{eqnarray}
\label{NovoAA1}\partial_x\mathcal{J}_x(x, \, k) &=& +
(-1)^{a}k^{b-1}\frac{\beta^{b}\alpha^{a}}{\Gamma(a)\Gamma(b)}\partial_\alpha^{a-1}\left\{\left[\alpha-2\sin\left(\frac{\alpha}{2}\right)e^{-k}\right]\exp(-\alpha x - \beta k)\right\}, \\
\label{NovoBB1}\partial_k\mathcal{J}_k(x, \, k) &=& -g\,(-1)^{b}\,x^{a-1}\frac{\beta^{b}\alpha^{a} }{\Gamma(a)\Gamma(b)}\partial_\beta^{b-1} 
\left\{\left[\beta-2\sin\left(\frac{\beta}{2}\right)e^{-x} \right]\, \exp(-\alpha x - \beta k)\right\}, \,\,
\end{eqnarray}
from which one has
\begin{eqnarray}
\label{NovoAA2}\mathcal{J}_x(x, \, k) &=& -
(-1)^{a}k^{b-1}\frac{\beta^{b}\alpha^{a}}{\Gamma(a)\Gamma(b)}\partial_\alpha^{a-1}\left\{\left[1-\frac{2}{\alpha}\sin\left(\frac{\alpha}{2}\right)e^{-k}\right]\exp(-\alpha x - \beta k)\right\}, \\
\label{NovoBB2}\mathcal{J}_k(x, \, k) &=& +g\,(-1)^{b}\,x^{a-1}\frac{\beta^{b}\alpha^{a} }{\Gamma(a)\Gamma(b)}\partial_\beta^{b-1} 
\left\{\left[1-\frac{2}{\beta}\sin\left(\frac{\beta}{2}\right)e^{-x} \right]\, \exp(-\alpha x - \beta k)\right\}. \,\,
\end{eqnarray}

\subsection{Wigner currents for modified LV systems -- Gaussian ensemble quantum features}

Analogously, for the Hamiltonian in the form of Eq.~\eqref{Original}, such that
\begin{eqnarray}
\label{t111Bss}
\partial_k^{2\eta+1}\mathcal{K}(k) &=& \sinh(k),\\
\label{t222Bss}
\partial_x^{2\eta+1}\mathcal{V}(x) &=& g \,\sinh(x), 
\end{eqnarray}
and straightforwardly from Eqs.~\eqref{imWA4}-\eqref{imWB4}, the following associated Wigner flow contributions are found for Gaussian ensembles (cf. Eq.~\eqref{gau})\footnote{With corresponding classical contributions given by
\begin{eqnarray}
\label{imWA4CC3mcl}
\partial_x\mathcal{J}^{\mathcal{C}\alpha}_x(x, \, k) &=& 2\alpha^2 \,x\,\sinh(k)\,
\mathcal{G}_{\alpha}(x, \, k),\\
\label{imWB4CC3mcl}
\partial_k\mathcal{J}^{\mathcal{C}\alpha}_k(x, \, k) &=& -2g\,\alpha^2 \,k\,\sinh(x)\,
\mathcal{G}_{\alpha}(x, \, k).
\end{eqnarray}},
\begin{eqnarray}
\label{imWA4CC3m}
\partial_x\mathcal{J}^{\alpha}_x(x, \, k) &=& 2 \sinh(k)\,\sin\left(\alpha^2\,x\right)\,e^{\frac{\alpha^2}{4}}\,\mathcal{G}_{\alpha}(x, \, k),\\
\label{imWB4CC3m}
\partial_k\mathcal{J}^{\alpha}_k(x, \, k) &=& -2 g\,\sinh(x)\sin\left(\alpha^2\,k\right)\,e^{\frac{\alpha^2}{4}}
\,\mathcal{G}_{\alpha}(x, \, k),
\end{eqnarray}
from which, after position or momentum integrations, one has
\begin{eqnarray}
\label{imWA4CCD4}\mathcal{J}^{\alpha}_x(x, \, k) &=& 
-\frac{i\,\alpha}{2\sqrt{\pi}} \,\sinh(k)\,e^{-\alpha^2 k^2}
\left\{\mbox{\sc{Erf}}\left[\alpha(x-i/2)\right]-\mbox{\sc{Erf}}\left[\alpha(x+i/2)\right]\right\},\\
\label{imWB4CCD4}\mathcal{J}^{\alpha}_k(x, \, k) &=& 
+\frac{i\,g\,\alpha}{2\sqrt{\pi}} \,\sinh(x)\,e^{-\alpha^2 x^2}
\left\{\mbox{\sc{Erf}}\left[\alpha(k-i/2)\right]-\mbox{\sc{Erf}}\left[\alpha(k+i/2)\right]\right\}.\,\,\,\,
\end{eqnarray}

\subsection{Wigner currents for modified LV systems -- Gamma distribution ensemble quantum features} 

Finally, for gamma distribution ensembles (cf. Eq.~\eqref{gaa}), from Eqs.~\eqref{t111Bss} and \eqref{t222Bss} once substituted into Eqs.~\eqref{imWA2bn} and \eqref{imWB2bn}, the following associated Wigner flow contributions\footnote{With corresponding classical contributions given by
\begin{eqnarray}
\label{NovoAA3cl}\partial_x\mathcal{J}^{\mathcal{C}}_x(x, \, k) &=& +\sinh(k)
(-1)^{a}k^{b-1}\frac{\beta^{b}\alpha^{a}}{\Gamma(a)\Gamma(b)}\partial_\alpha^{a-1}\left\{\alpha\exp(-\alpha x - \beta k)\right\}, \\
\label{NovoBB3cl}\partial_k\mathcal{J}^{\mathcal{C}}_k(x, \, k) &=& -g \sinh(x)\,(-1)^{b}\,x^{a-1}\frac{\beta^{b}\alpha^{a} }{\Gamma(a)\Gamma(b)}\partial_\beta^{b-1} 
\left\{\beta\, \exp(-\alpha x - \beta k)\right\}. \,\,
\end{eqnarray}},
 are gotten,
\begin{eqnarray}
\label{NovoAA3}\partial_x\mathcal{J}_x(x, \, k) &=& +2\sinh(k)\,(-1)^{a}k^{b-1}\frac{\beta^{b}\alpha^{a}}{\Gamma(a)\Gamma(b)}\partial_\alpha^{a-1}\left[\sin\left(\frac{\alpha}{2}\right)\exp(-\alpha x - \beta k)\right], \\
\label{NovoBB3}\partial_k\mathcal{J}_k(x, \, k) &=& -2g\,\sinh(x)\,(-1)^{b}\,x^{a-1}\frac{\beta^{b}\alpha^{a} }{\Gamma(a)\Gamma(b)}\partial_\beta^{b-1} 
\left[\sin\left(\frac{\beta}{2}\right) \, \exp(-\alpha x - \beta k)\right],\,\,
\end{eqnarray}
from which one has
\begin{eqnarray}
\label{NovoAA4}\mathcal{J}_x(x, \, k) &=& -\sinh(k)\,(-1)^{a}k^{b-1}\frac{\beta^{b}\alpha^{a}}{\Gamma(a)\Gamma(b)}\partial_\alpha^{a-1}\left[\frac{2}{\alpha}\sin\left(\frac{\alpha}{2}\right)\exp(-\alpha x - \beta k)\right], \\
\label{NovoBB4}\mathcal{J}_k(x, \, k) &=& +g\,\sinh(x)\,(-1)^{b}\,x^{a-1}\frac{\beta^{b}\alpha^{a} }{\Gamma(a)\Gamma(b)}\partial_\beta^{b-1} 
\left[\frac{2}{\beta}\sin\left(\frac{\beta}{2}\right) \, \exp(-\alpha x - \beta k)\right]. \,\,
\end{eqnarray}

From the results for all the above cases, stationarity and Liouvillianity quantifiers, $\mbox{\boldmath $\nabla$}_{\xi}\cdot\mbox{\boldmath $\mathcal{J}$}$ and $\mbox{\boldmath $\nabla$}_{\xi} \cdot \mathbf{w}$, can be straightforwardly computed (with $\mbox{\boldmath $\nabla$}_{\xi} \cdot \mathbf{w}$ as from Eq.~\eqref{altdiv2}).

\begin{figure}
\includegraphics[scale=0.16]{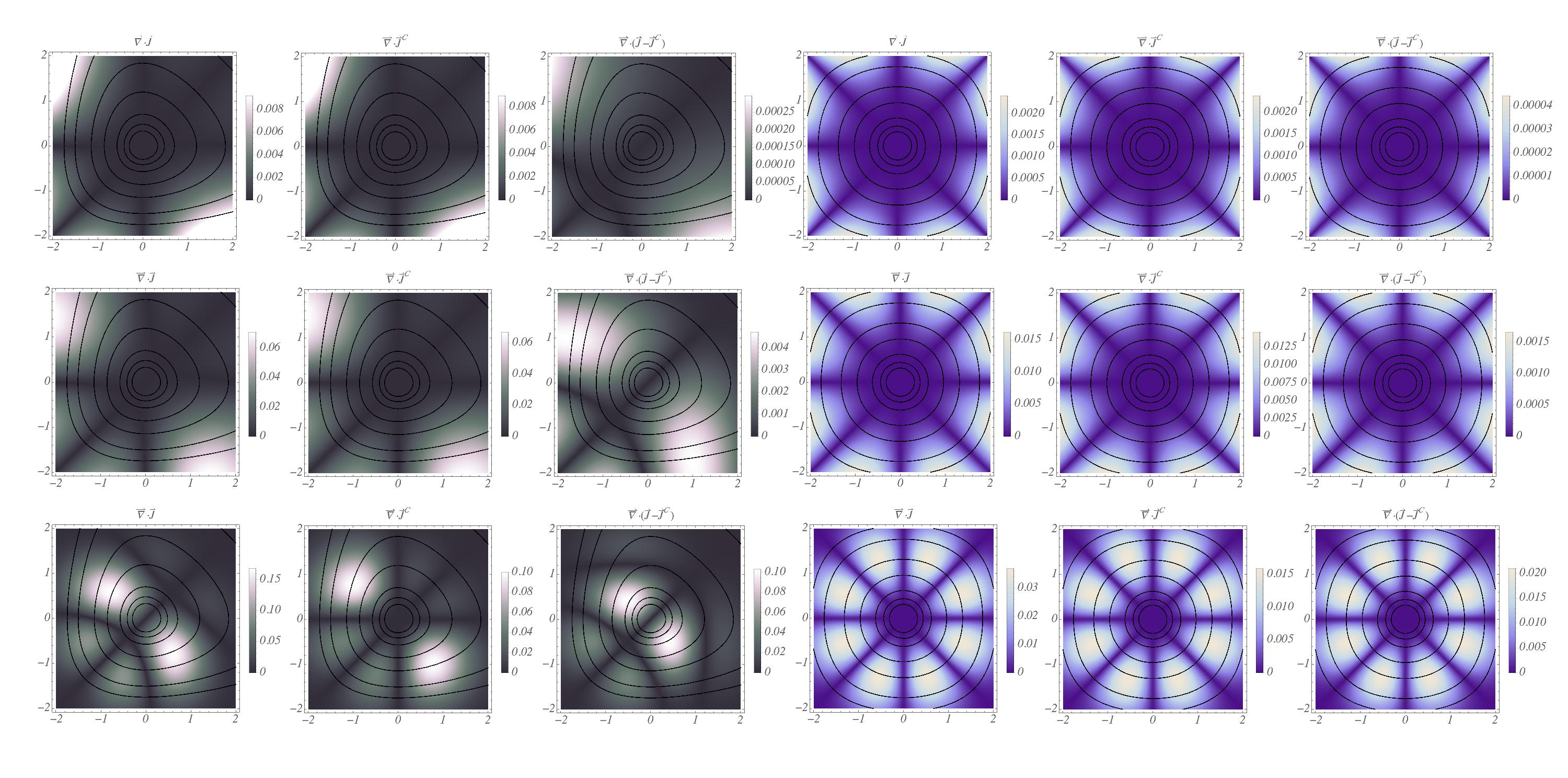}
\renewcommand{\baselinestretch}{.6}
\vspace{-1.2 cm}\caption{\footnotesize
(Color online)
Absolute values of stationarity quantifier, $\mbox{\boldmath $\nabla$}_{\xi}\cdot\mbox{\boldmath $\mathcal{J}$}^{\alpha}$ ({\em first and forth columns}) from typically classical, $\mbox{\boldmath $\nabla$}_{\xi}\cdot\mbox{\boldmath $\mathcal{J}$}^{\mathcal{C}\alpha}$ ({\em second and fifth columns}), and quantum, $\mbox{\boldmath $\nabla$}_{\xi}\cdot(\mbox{\boldmath $\mathcal{J}$}^{\alpha}-\mbox{\boldmath $\mathcal{J}$}^{\mathcal{C}\alpha})$ ({\em third and sixth columns}) contributions for Wigner currents convolved by Gaussian ensembles. Results are depicted through the light-dark background color scheme, with darker regions approaching stationarity for typical (first three columns) and modified (last three columns) prey-predator-like dynamics. Classical trajectories are shown as a collection of black lines.
The results are for the increasing spreading characteristic of the Gaussian function, from $\alpha =1/4$ (first row), $1/2$ (second row) and $1$ (third row).}
\label{Novass1}
\end{figure}

\begin{figure}
\includegraphics[scale=0.16]{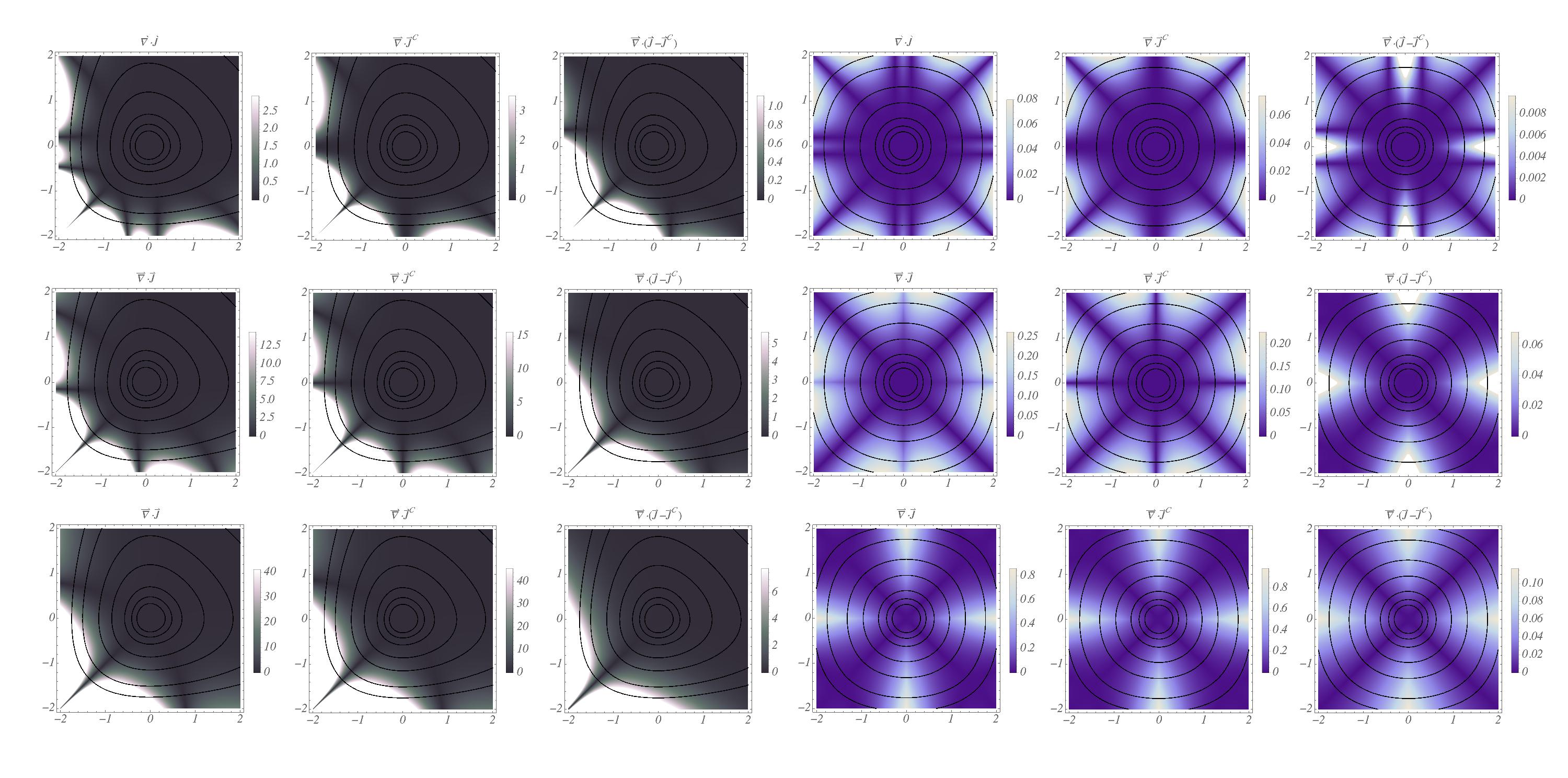}
\renewcommand{\baselinestretch}{.6}
\vspace{-1.2 cm}\caption{\footnotesize
(Color online)
Absolute values of stationarity quantifier, $\mbox{\boldmath $\nabla$}_{\xi}\cdot\mbox{\boldmath $\mathcal{J}$}^{\alpha}$ ({\em first and forth columns}) from typically classical, $\mbox{\boldmath $\nabla$}_{\xi}\cdot\mbox{\boldmath $\mathcal{J}$}^{\mathcal{C}\alpha}$ ({\em second and fifth columns}), and quantum, $\mbox{\boldmath $\nabla$}_{\xi}\cdot(\mbox{\boldmath $\mathcal{J}$}^{\alpha}-\mbox{\boldmath $\mathcal{J}$}^{\mathcal{C}\alpha})$ ({\em third and sixth columns}) contributions for Wigner currents convolved by gamma (black color scheme) and Laplacian (purple color scheme) distribution ensembles.
Results are depicted through the same light-dark background color scheme from Fig.~\ref{Novass1} for typical (first three columns) and modified (last three columns) prey-predator-like dynamics.
The parameters of the gamma/Laplacian distribution are chosen as $\alpha=\beta=1$ and $a=b=2$ (first row), $3$ (second row) and $4$ (third row).}
\label{Novass2}
\end{figure}

Results for the stationarity quantifiers, for classical and quantum regimes, are depicted in Figs.~\ref{Novass1} and \ref{Novass2} for Gaussian and gamma/Laplacian distribution ensembles, respectively.
For highly spread Gaussian distributions, the pattern of stationarity is typically from classical origin for both, typical and modified, prey-predator-like dynamics, with quantum modifications just introducing tiny fluctuations which do not affect the distribution of stationary states in the phase space. Peaked Gaussian distributions (increasing $\alpha$) localizes and intensifies the quantum distortions, destroying the stationarity of the orbits closer to the origin ($x=k=0$) which, classically, would approach the harmonic ones. Due to the higher level of symmetry of the modified LV Hamiltonian, when compared with the typical one, in the distribution of quantum states over the phase space, the typical prey-predator-dynamics is more sensible to the convolution with highly peaked Gaussian distribution. For the modified LV Hamiltonian configuration, phase-space orbits highly closed to the origin have their associated stationarity just slightly affected .

For Wigner currents convolved by gamma distributions, as depicted in the first set of plots from Fig.~\ref{Novass2}, for the typical LV Hamiltonian configuration, a relative comparison is not convenient since the gamma distribution is not parity symmetric. Moreover, gamma/Laplacian distributions have four parameters that can be arbitrarily manipulated. One just notices that stationarity is more spreadably distributed in this case.
For Wigner currents convolved by Laplacian distributions, which is possible due to the higher level of symmetry of the modified LV Hamiltonian in the phase-space plane, as depicted in the second set of plots from Fig.~\ref{Novass2}, stationarity is also more spreadably distributed than for the Gaussian convolution, with quantum fluctuations affecting the stationarity of phase-space states around either $x=0$ or $k=0$ coordinates, but far from the origin ($x=k=0$).

Due to the higher number of involved parameters, our partial conclusion is that stationarity, even depending on both classical and quantum contributions, is highly manipulable according to the statistical distribution driving the quantum fluctuations.

\begin{figure}
\vspace{-1.2 cm}\includegraphics[scale=0.24]{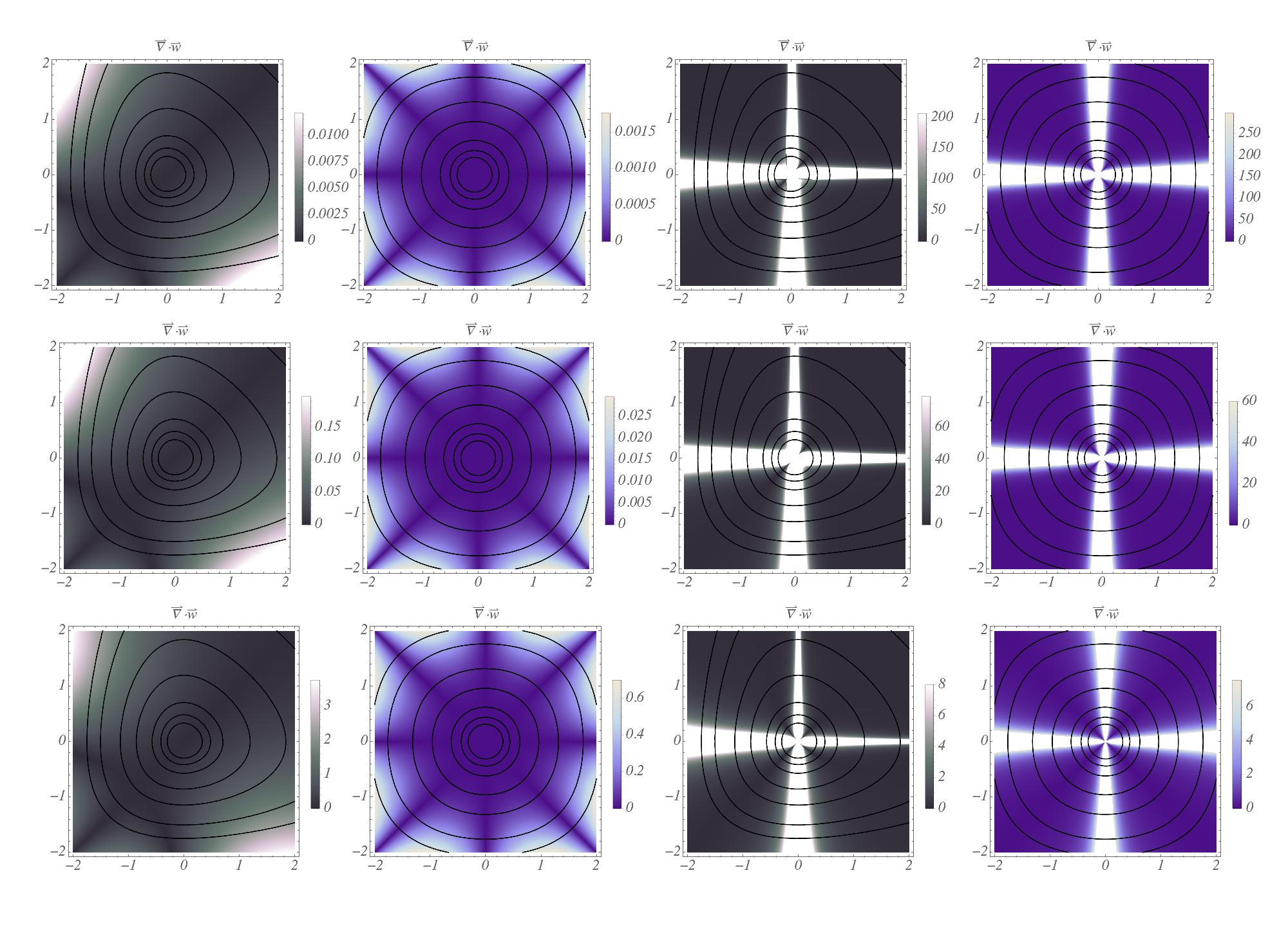}
\renewcommand{\baselinestretch}{.6}
\vspace{-1.2 cm}\caption{\footnotesize
(Color online)
Absolute values of Liovillianity quantifier, $\mbox{\boldmath $\nabla$}_{\xi} \cdot \mathbf{w}$, for typical ({\em first and third columns}) and modified ({\em second and forth columns}) LV Hamiltonians, for Wigner currents convolved by Gaussian ({\em first and second columns}), gamma ({\em third column}) and Laplacian ({\em forth column}) distribution ensembles.
Results are depicted through the same light-dark (blue) background color scheme and the same parameters from Figs.~\ref{Novass1} and \ref{Novass2}.}
\label{Novass3}
\end{figure}

On the other hand, as discussed in the previous sections, the non-Liouvillianity has uniquely a quantum origin.
Fig.~\ref{Novass3} depicts the result for Gaussian and gamma/Laplacian distribution ensembles for both typical and modified
LV Hamiltonians.
Qualitatively, from the preliminary perspective for the results from Fig.~\ref{Novass3}, the quantum patterns are not affected by the magnitude of either Gaussian ($\alpha$) or gamma/Laplacian ($\alpha$, $\beta$, $a$ and $b$) parameters which just modulate the quantum corrections. Decoupled from stationarity effects, Gaussian and gamma/Laplacian distributions produce opposite patterns with respect to the quantum fluctuations.
Therefore, the choice of the statistical distribution in driving the quantum fluctuations, besides the appeal related to their analytical manipulability when convolved with Wigner current results, has a discriminant role in quantifying the non-Liouvillianity.

\section{Conclusions}

An extension of the phase-space Weyl-Wigner QM to the subset of Hamiltonians in the form of $H(q,\,p) = {K}(p) + {V}(q)$, which includes out-of-the-ordinary contributions for the kinetic term, was analytically investigated through their Wigner flow properties.
Generalized Liouvillian and stationary properties were obtained for Gaussian and gamma (Laplacian) distribution quantum ensembles in order to account for the exact pattern of quantum fluctuations over a classical phase-space scenario since the overall quantum distortion was obtained in terms of convergent infinite series expansions over $\hbar^{2}$.
General results were then specialized to some Hamiltonians which encompass, for instance, the investigation of prey-predator-like quantum problems.

A preliminary analysis of the classical pattern of typical and modified prey-predator-like Hamiltonian systems shows that semi-classical quantitative aspects can be straightforwardly obtained from LV and modified LV equations. These include the identification of a Bohr-Sommerfeld quantization rule for typical LV systems, and the observation of persistent prey and predator dominating regimes for modified LV systems. 
Through Wigner flow operators describing fluctuations over stationary and Liovillian regimes, information flow aspects as well as quantum-like deviations from the typical prey-predator classical system were then quantified.
The analysis included Hamiltonians, $H(q,\,p) = {K}(p) + {V}(q)$, with momentum contributions given by $K(p) = p/p_0 + \exp(-p/p_0)\equiv\mathcal{K}(k) = k+e^{ -k}$ and $K(p) \propto \cosh(p/p_0)\equiv \mathcal{K}(k) = \cosh(k)$, where the persistence of prey and predator domains could be identified for the latter one.

In fact, considering that the hyperbolic behavior is dominant in the description of the competitive ecological equilibrium of populations, the Hamiltonian formulation of LV-like systems \cite{LV1,Novo21B2,Novo222}, for instance, when applied in the investigation of stochastic systems \cite{Allen,Grasman}, matches the dynamical equations arising from a phase-space formulation whose associated trajectories could be extended to the non-commutative context, $[x,\,k]= i$, through the WW formalism. 
Hence, exploring some issues related to the phase-space dynamics, which includes the systematic evaluation of quantum effects, and their consequences as basic features of classical-quantum emergence, further supports our analysis of more realistic non-linear models which can also admit scenarios subjected to statistical constraints.

That can be the case of considering gamma distributions, for instance, for modeling phenomena composed by several sub-events
which occurs, for instance, either in a sequence of cell-division events \cite{26} or through compensatory mutations for a given mutation \cite{27}.
Of course, from the perspective of adopting a quantum approach, cell division and mutation are immensely complex mechanisms. In fact, one would not expect straightforward results from quantum mechanical calculations \cite{Bord13,Bord19}. Elementary approximations, however, may capture the relevant information about microscopical biological systems in order to incorporate some characteristic features of the quantum-theoretical description \cite{Bord13}.
As an example, directed adaptive mutation mediated by quantum mechanical effects have been considered in the investigation of carcinogenesis \cite{Bord19}.
More related to the Wigner seminal proposal on the quantum correction for thermodynamic ensembles \cite{Wigner,Novo21A,Novo21B}, 
the determination of the tunneling rate associated to spontaneous point mutations in DNA \cite{DNA1} have been investigated.
Through an open quantum system description for the quantum tunneling contribution to the proton transfer rate in DNA \cite{DNA2}, the equivalent phase-space formulation of the Caldeira and Leggett master equation \cite{Caldeira}, the so-called Wigner-Moyal-Caldeira-Leggett equation (cf. Eq.~(1) from Ref.~\cite{DNA2}), is the setup for the calculations. Therefore, the Wigner framework, in the above context, can be the bridge between this kind of microscopic systems and QM.

Of course, our results neither exhaust the still nascent algorithms required for computing quantum effects in such kind of non-linear systems nor is in the ultimate form for being connected with such a biological phenomenology, but they are, in our view, a relevant step forward.

{\em Acknowledgments -- The work of AEB is supported by Grant No. 2023/00392-8 (S\~ao Paulo Research Foundation (FAPESP)) and Grant No. 301485/2022-4 (CNPq).}


\begin{thebibliography}{99}

\bibitem{Wigner}
E. Wigner, Phys. Rev. {\bf 40} 749 (1932).
\bibitem{Case}
W. B. Case, Am. J. Phys. {\bf 76}, 937 (2008).
\bibitem{Weyl}
H. Weyl, Zeitschrift f\"ur Physik {\bf 46}, 1 (1927).
\bibitem{Abr65}
A. A. Abrikosov, L. P. Gorkov and I. E. Dzyaloshinskii, {\em Quantum Field Theoretical Methods in Statistical Physics}, 2nd Ed. (Pergamon, Oxford 1965).
\bibitem{Sch81}
L. S. Schulman, {\em Techniques and Applications of Path Integral} (Jonh Wiley \& Sons Inc., New York 1981).
\bibitem{Par88}
G. Parisi, {\em Statistical Field Theory} (Benjamin/Cummings, New York 1988).
\bibitem{Zurek02}
W. H. Zurek, Phys. Today {\bf 44}, 36 (1991).
\bibitem{Bernardini13A}
A. E. Bernardini and O. Bertolami, Phys. Rev. A {\bf 88}, 012101 (2013).
\bibitem{Leal2019}
P. Leal, A. E. Bernardini and O. Bertolami, J. Phys. A {\bf 72}, 375302 (2019).
\bibitem{Catarina}
C. Bastos, O. Bertolami, N. C. Dias and J. N. Prata, J. Math. Phys. {\bf 49}, 072101 (2008). 
\bibitem{Bernardini13B}
C. Bastos, A. E. Bernardini, O. Bertolami, N. C. Dias, J. N. Prata, Phys. Rev. D {\bf 88}, 085013 (2013).
\bibitem{PhysicaA}
J. F. G. Santos, A. E. Bernardini and C. Bastos, Phys. A: Stat. Mech. Appl.{\bf 438}, 340 (2015).
\bibitem{Catarina001}
C. Bastos, O. Bertolami, N. C. Dias and J. N. Prata, Int. J. Mod. Phys. A {\bf 24}, 2741 (2009). 
\bibitem{Stein}
 L. A. Rozema, A. Darabi, D. H. Mahler, A. Hayat, Y. Soudagar, and A. M. Steinberg, Phys. Rev. Lett. {\bf 109}, 100404 (2012).
\bibitem{Bernardini13C}
C. Bastos, A. E. Bernardini, O. Bertolami, N. C. Dias, J. N. Prata, Phys. Rev. D {\bf 90}, 045023 (2014).
\bibitem{Bernardini13E}
C. Bastos, A. E. Bernardini, O. Bertolami, N. C. Dias, J. N. Prata, Phys. Rev. D {\bf 91}, 065036 (2015).
\bibitem{JCAP18}
A. E. Bernardini, P. Leal and O. Bertolami, JCAP {\bf 2018}(02), 025 (2018).
\bibitem{NossoPaper}
A. E. Bernardini and O. Bertolami, EPL {\bf 120}, 20002 (2017); A. E. Bernardini and O. Bertolami, Journal of Physics: Conf. Series {\bf 1275}, 012032 (2019).
\bibitem{Meu2018}
A. E. Bernardini, Phys. Rev. A {\bf 98}, 052128 (2018).
\bibitem{Bernardini2020-02}
C. Fernando e Silva and A. E. Bernardini, Phys. A: Stat. Mech. Appl. {\bf 558}, 124915 (2020).
\bibitem{Steuernagel3}
O. Steuernagel, D. Kakofengitis and G. Ritter, Phys. Rev. Lett. {\bf 110}, 030401 (2013).
\bibitem{Novo21A}
A. E. Bernardini and O. Bertolami, Phys. Rev. A {\bf 105}, 032207 (2022).
\bibitem{Novo21B}
A. E. Bernardini and O. Bertolami, Phys. Rev. E {\bf 106}, 024202 (2022). 
\bibitem{Novo21B2}
A. E. Bernardini and O. Bertolami, Phys. Rev. E {\bf 107}, 044201 (2023).
\bibitem{0001}
S. H. V. Leme de Mattos, J. R. C. Piqueira, J. Vasconcelos-Neto and F. M. Orsatti, {\em Measuring Q-Bits in Three Trophic Level Systems}, Ecological Modelling {\bf 200}, 183 (2007).
\bibitem{0002}
E. Del Giudice, R. M. Pulselli and E. Tiezzi, {\em Thermodynamics of Irreversible Processes and Quantum Field Theory: An Interplay for the Understanding of Ecosystem Dynamics}, Ecological Modelling {\bf 220}, 1874 (2009).
\bibitem{0003}
A. D. Kirwan Jr.,{\em Quantum and Ecosystem Entropies}, Entropy {\bf 10}, 58 (2008).
\bibitem{0004}
S. E. Jorgensen and E. Tiezzi, {\em Preface to Workshop on ``Emergence of Novelties'', 9--16 October 2008, Pacina, Siena, Italy}, Ecological Modelling {\bf 220}, 1855 (2009).
\bibitem{Novo222}
A. E. Bernardini and O. Bertolami, Foundations of Physics {\bf 53}, 63 (2023).
\bibitem{PP00}
D. P. Paula {\it et al.}, {\em Detection and decay rates of prey and prey symbionts in the gut of a predator through metagenomics},
Molecular Ecology Resources {\bf 15}, 880 (2015).
\bibitem{PP01}
T. Fujii and T. Rondelez, {\em Predator-prey molecular ecosystems} ACS Nano {\bf 7}, 27 (2013).
\bibitem{PP02}
I. R. Epstein and J. A. Pojman, {\em An Introduction to Nonlinear Chemical Dynamics} (Oxford University Press, New York 1998).
\bibitem{PP03}
J. Ackermann, B. Wlotzka and J. S. McCaskill, {\em In Vitro DNA-Based Predator Prey System with Oscillatory Kinetics},
Bull. Math. Biol. {\bf 60}, 329 (1998).
\bibitem{PP04}
B. Wlotzka and J. S. McCaskill, {\em A Molecular Predator and Its Prey: Coupled Isothermal Amplification of Nucleic Acids},
Chem. Biol. {\bf 4}, 25 (1997).
\bibitem{LV1}
A. J. Lotka, {\em Elements of physical biology} (Williams \& Wilkins Co., Baltimore 1925).
\bibitem{LV2}
V. Volterra, {\em Variazioni e fluttuazioni del numero d'individui in specie animali conviventi}, Mem. R. Accad. Naz. Lincei. (Ser. VI) 2, 31-113 (1926).
\bibitem{Ballentine}
L. E. Ballentine, {\em Quantum Mechanics: a Modern Development}, pp. 633 (World Scientific, 1998).
\bibitem{26}
A. Golubev, J. Theo. Biology {\bf 393}, 203 (2016).
\bibitem{27}
A. Poon, B. H. Davis and L. Chao, Genetics {\bf 170}, 1323 (2005).
\bibitem{PRE-LV}
M. Parker and A. Kamenev, {\em Extinction in the Lotka-Volterra model}, Phys. Rev. E {\bf 80}, 021129 (2009).
\bibitem{SciRep02}
T. Tahara {\em et al.}, {\em Asymptotic stability of a modified Lotka-Volterra model with small immigrations}, Sci Rep {\bf 8}, 7029 (2018).
\bibitem{PRE-LV2}
N. R. Smith and B. Meerson, {\em Extinction of oscillating populations}, Phys. Rev E {\bf 93}, 032109 (2016).
\bibitem{RPSA-LV}
Yi-An Ma and Hong Qian, {\em A thermodynamic theory of ecology: Helmholtz theorem for Lotka-Volterra equation, extended conservation law, and stochastic predator-prey dynamics}, Proceedings of the Royal Society A {\bf 471}, 20150456 (2015).
\bibitem{Anna}
A. Vanselow, S. Wieczorek and U. Feudel, {\em When very slow is too fast -- collapse of a predator-prey system}, Journal of Theoretical Biology {\bf 479}, 64 (2019).
\bibitem{Nature01}
M. Kumar, B. Ji, K. Zengler and J. Nielsen, {\em Modelling approaches for studying the microbiome}, Nat. Microbiol. {\bf 4}, 1253 (2019).
\bibitem{Nature02}
S. Butler and J. P. O'Dwyer, {\em Stability criteria for complex microbial communities}, Nat. Commun. {\bf 9}, 2970 (2018).
\bibitem{PRE-LV3}
G. Szabo and T. Czaran, {\em Phase transition in a spatial Lotka-Volterra model}, Phys. Rev. E {\bf 63}, 061904 (2001).
\bibitem{Grasman}
J. Grasman and O. A. van Herwaarden, {\em Asymptotic methods for the Fokker-Planck equation and the exit problem in applications} (Springer, Berlin 1999).
\bibitem{Allen}
L. J. S. Allen, {\em An introduction to stochastic processes with applications to biology}- 2nd Ed.(Chapman \& Hall-CRC, New York 2010).
\bibitem{Neumann}
J. von Neumann, {\em Mathematical Foundations of Quantum Mechanics}, Translated by R. T. Beyer, (Princeton University Press, 1955).
\bibitem{Bord13}
M. Bordonaro and V. Ogryzko, Biosystems {\bf 112}, 11 (2013).
\bibitem{Bord19}
M. Bordonaro, Biosystems {\bf 178}, 16 (2019).
\bibitem{DNA1}
M. Winokan, L. Slocombe, J. Al-Khalili and M. Sacchi, Scientific Reports {\bf 13}, 21749 (2023).
\bibitem{DNA2}
L. Slocombe, M. Sacchi and J. Al-Khalili, Communications Physics {\bf 5}, 109 (2022).
\bibitem{Caldeira}
A. O. Caldeira and A. J. Leggett, Phys. A: Stat. Mech. Appl. {\bf 121}, 587 (1983).
\end{thebibliography}
\end{document}